\documentclass[paper,notoc]{JHEP3}

\author{L.A. Harland-Lang$^1$, V.A. Khoze$^{2,3}$, M.G. Ryskin$^{2,4}$, W.J. Stirling$^{1,2}$\\ 
  $^1$Cavendish Laboratory, University of Cambridge,
  J.J.\ Thomson Avenue, Cambridge, CB3 0HE, UK\\
  $^2$ Department of Physics and Institute for Particle Physics Phenomenology, University of Durham, DH1 3LE, UK\\
$^3$ School of Physics \& Astronomy, University of Manchester,
Manchester M13 9PL, UK
$^4$ Petersburg Nuclear Physics Institute, Gatchina, St. Petersburg, 188300, Russia}

\title{Standard Candle Central Exclusive Processes at the Tevatron and LHC}

\abstract{Central exclusive production (CEP) processes in high-energy proton -- (anti)proton collisions 
offer a very promising framework within which to study both novel aspects of QCD and new physics signals. Among the many interesting
processes that can be studied in this way, those involving the production of heavy $(c,b)$ quarkonia and $\gamma\gamma$ states 
have sufficiently well understood theoretical properties and sufficiently large cross sections that they can serve  
as `standard candle' processes with which we can benchmark predictions for new physics CEP at the CERN Large Hadron Collider. 
Motivated by the broad agreement with theoretical predictions of recent CEP measurements at the Fermilab Tevatron, we perform a
detailed quantitative study of heavy quarkonia ($\chi$ and $\eta$) and $\gamma\gamma$ production at the Tevatron, RHIC and LHC, 
paying particular attention to the various uncertainties in the calculations. Our results confirm the rich phenomenology that 
these production processes offer at present and future high-energy colliders.}

\preprint{IPPP/10/32\\ DCPT/10/64 \\ Cavendish-HEP-10/08}

\usepackage{multirow}
\usepackage{helvet}
\usepackage{amsmath}
\usepackage{amssymb}
\usepackage{setspace}
\usepackage{setspace}
\usepackage[dvips]{graphicx}
\usepackage{epsfig}
\def\lesim{ \;\raisebox{-.7ex}{$\stackrel{\textstyle <}{\sim}$}\; }
\def\be{\begin{equation}}
\def\ee{\end{equation}}

\begin{document}

\section{Introduction}

Recently there has been a renewal of interest in studies of central exclusive production (CEP) processes in high-energy proton -- (anti)proton collisions, see \cite{acf}~--~\cite{Pasechnik:2009qc}. In particular, such measurements represent a very promising way to study the properties of new particles, from exotic hadrons to the Higgs boson, see for example \cite{acf},\cite{DR}~-~\cite{Klempt}. The CEP of an object $X$ may be written in the form 
\begin{equation}\nonumber
pp({\bar p}) \to p+X+p({\bar p})\;,
\end{equation}
where $+$ signs are used to denote the presence of large rapidity gaps. An attractive advantage of these reactions is that they provide an especially clean environment in which to measure the nature and quantum numbers (in particular, the spin and parity) of new states, see for example \cite{Kaidalov03}, \cite{CK}~-~\cite {BSM}. An important example is the CEP of the Higgs boson~\cite{acf,epip}, \cite{Khoze00}~-~\cite{FP420}.
 This  provides a novel route to study in detail the Higgs sector at the LHC, which is complementary to the conventional measurements, and
gives a strong motivation for the addition of near-beam proton detectors to enhance the discovery and physics potential of the ATLAS and CMS detectors at the LHC \cite{FP420}~-~\cite{royon}.

While the CEP mechanism undoubtedly allows a promising framework within which to study new physics signals at the LHC, it can also be used to study Standard Model physics. In particular, we should expect conventional lighter mass states, such as $c\overline{c}$ and $b\overline{b}$ quarkonia, diphotons ($\gamma\gamma$) and dijets ($jj$), to be produced via the same CEP mechanism as Higgs bosons (or other new colourless objects) at the LHC, but with much larger cross sections. These processes are not only of interest in their own right, but can also serve as an important check on the CEP theoretical framework, and may help to reduce some of the uncertainties involved in the model, see for instance \cite{early,epip,HarlandLang09,KMRprosp}.
 Crucially, central exclusive $\gamma\gamma$ \cite{CDFgg}, dijet \cite{CDFjj} and $\chi_c$ \cite{Aaltonen09} production have indeed been successfully observed by the CDF Collaboration at the Tevatron.\footnote{Recently the D0 collaboration at the Tevatron has reported evidence
for CEP of high-mass dijets \cite{Rangel}, and more CDF exclusive data on $\gamma\gamma$ production will be available in the very near future~\cite{Albrow}.}
 These can therefore serve as standard candle processes with which we can check our predictions for new physics CEP at the LHC by measurements made at the Tevatron and in the early stages of LHC running. Indeed, the observed rates of all three CEP processes measured at the Tevatron are in broad agreement with theoretical expectations \cite{HarlandLang09,Khoze00a,Khoze04,KMRprosp,Khoze04gg}, which lends credence to the predictions for exclusive Higgs production at the LHC. 

Among the potential standard candle processes, the CEP of heavy quarkonia ($\chi_{(c,b)}$ and $\eta_{(c,b)}$) states plays a special role \cite{HarlandLang09,Khoze04} (see also \cite{teryaev,Pasechnik:2009qc}, \cite{Pump}~--~\cite{RPtheor}). First, as is well known, heavy quarkonium production provides  a valuable tool to test the ideas and methods of the QCD physics of bound states, such as  effective field theories, lattice QCD, NRQCD, etc. (see, for example, \cite{Bodwin}~--~\cite{simon} for theoretical reviews). Second, heavy quarkonium CEP exhibits characteristic features, based on Regge theory, that depend on the particle spin and parity $J^P$, and these are altered by both the loop integration around the internal gluon momentum $Q_\perp$ and non-zero outgoing proton $p_\perp$ effects as well as by screening corrections arising from multi-Pomeron exchanges. Measurement of these effects, in particular the distributions of the outgoing proton momenta \cite{KMRtag}, would provide a valuable source of spin-parity information about the centrally produced system as well as constituting an important test of the overall theoretical formalism, as we will describe in Section~\ref{surveff}.

It is worthwhile to recall that the reconstruction of the bottomonium spectroscopy is still incomplete and, despite a good deal of valuable information
on the $b\overline{b}$ states and transitions, various issues  remain so far unresolved. Although the $\Upsilon({}^3S_{1})$ state was discovered in 1977 \cite{herb}, its spin-singlet partner  $\eta_{b}({}^1S_{0})$ was found  more than thirty years later \cite{eta}\footnote{The study of $\eta$ CEP could provide additional information about the dynamics of heavy quarkonia since, for example, within the non-relativistic potential model the $\eta$ production amplitude is related to the $Q\bar{Q}$ wavefunction at the origin, rather than to its derivative as in the case of the $\chi$ states. We recall that the first data on the hyperfine splitting between the $\eta_b$ and $\Upsilon(1S)$ states and the branching fraction of this M1 transition have already allowed a critical test of the various QCD and potential model predictions.}, while the spin assignment of the $P$-wave states $\chi_{bJ}$ still needs experimental confirmation \cite{PDG}. The central exclusive production mechanism, with its spin-parity analyzing capability, could therefore potentially provide a way to establish the spin-parity assignment of the $C$-even  $b\overline{b}$ states. 

Finally, we note that in recent years a whole zoology of exotic hadrons, in particular new charmonium-like states, have been observed both at B-factories and at the Tevatron, see for example \cite{exot}. There exist a variety of theoretical interpretations of these new particles,
 although in many cases their spin-parity assignment remains undetermined. The CEP process has the promising potential to resolve this issue, and in doing so could shed more light on the dynamical origin of these new objects.

The case of $\gamma\gamma$ CEP is also of much interest,  allowing a more detailed probing of the underlying theory. Specifically, we can measure the distribution of the cross section in $M_{\gamma\gamma}$ (by changing the cut on the photon $E_\perp$) and compare this with the theoretical predictions, in a similar manner to the analysis performed for exclusive dijet production~\cite{CDFjj}, where good agreement between theory and experiment was found, see \cite{epip,frascati}. This can also allow a test of the theory at higher $M_{\gamma\gamma}$ scales than in the case of $\chi_c$ CEP, where the low mass scale leads to large uncertainties in the theoretical predictions. An exclusive $\gamma\gamma$ search has been performed by CDF \cite{CDFgg}, allowing such a comparison to be made. Moreover, if the (existing and forthcoming) CDF data are combined with additional measurements of $\gamma\gamma$ CEP at the LHC, this would also provide the possibility of a more detailed study of the predicted energy dependence of the cross section, which we note will
be controlled in a non-trivial way by the energy dependence of soft rescattering effects (both `enhanced' and `eikonal') as well as the gluon density $x$ dependence. We recall that the theoretically most challenging contribution to the `enhanced' absorption effects, which break soft-hard factorization, depends mainly on the size of the rapidity gap $\sim \ln (s/M^2)$ of the exclusive process, see \cite{epip,Ryskin09}\footnote{For the most recent publication on enhanced rescattering and references see \cite{ostap}.}. In the case of $\chi_c$ CEP at the Tevatron this quantity exceeds the gap size for Higgs CEP at the LHC, while for exclusive $\gamma\gamma$ production at the Tevatron the corresponding gap size is similar to the LHC Higgs case. The observation of $\chi_b$ and $\gamma \gamma$ CEP at the LHC would therefore provide a unique opportunity to probe the rapidity gap survival factor $S^{\rm LHC}_{\rm enh}$ at the record large values of $s/M^2$, but still comfortably in the perturbative domain. A further interesting comparison to make would be between the $\gamma\gamma$ (for $M_{\gamma\gamma}\approx 10$ GeV) and the $\chi_{b0}$ processes: taking the ratio of these cross sections, various uncertainties (in particular, that due to absorption effects) would cancel.

The observation of these standard candle processes at the LHC, in combination with the pre-existing CDF data, would therefore provide a very interesting source of information for a more detailed analysis of the theoretical framework of CEP. Motivated by these considerations, we  consider in this paper the CEP of an object $X$ at the Tevatron and LHC ($\sqrt{s}=7,10,14$ TeV), where $X=c\overline{c}$ or $b\overline{b}$ mesons, in particular $\chi_{q(0,1,2)}$ and $\eta_q$ states, where $q=c,b$, as well as $X=\gamma\gamma$. Observation of these processes at the LHC would help build upon the previous CDF observation of $\chi_c$ and $\gamma\gamma$ production, while in the case of $\eta_{c,b}$ and $\chi_b$ production, which we note are unlikely to be observed at the Tevatron, these would represent completely new observations which are certainly worth pursuing. 

We also note that $\chi_{c,b}$ CEP is a potential observable for future planned $pp$ collisions at RHIC, for which a physics programme with tagged forward protons, covering a range of c.m.s energy values up to 500 GeV, already exists. In particular, the existing pp2pp experimental setup at the STAR detector includes Roman Pot detectors with a low enough central system mass $M_X$ acceptance to observe central exclusive $\chi_c$ production with tagged forward protons (the installation of further forward detectors is planned for the near future, see for example~\cite{Guryn08,LeeDIS}). Significantly, this would in principle include a measurement of the proton $p_\perp$ and the azimuthal correlations between the forward protons~\cite{Guryn}, which would provide an important test of the overall theoretical framework (see Section~\ref{surveff}) as well as potentially shedding some light on the previous Tevatron $\chi_c$ CEP data. We therefore also provide a cross section estimate for  $\chi_c$ production at the benchmark collision energy value of $\sqrt{s}=500$ GeV, while leaving a fuller Monte Carlo treatment of the predicted forward proton $p_\perp$ and $\phi$ distributions to a forthcoming publication~\cite{HKRSrhic}.

However, it should be noted that while these lower mass CEP processes have much higher cross sections and are, therefore, experimentally more accessible, theoretically they are quite challenging. In particular, as the mass $M_X$ of the central system $X$ is reduced, the uncertainties present in the calculation tend to increase. Specifically, as has been noted in the case of $\chi_c$ CEP~\cite{HarlandLang09,Khoze04}, the application of the perturbative CEP formalism is not completely valid at such low $M_X$ scales, where the process is not truly perturbative. Moreover, as we shall discuss in Section~\ref{gamres}, at the low $x$ and $Q^2$ values relevant to these calculations, there is quite a large uncertainty in the available PDF sets, which can act as a significant source of uncertainty in the cross section predictions, in particular at LHC energies. When we consider that there is also the uncertainty in the soft survival effects (and their energy dependence) to include, then clearly any theoretical predictions should be treated as estimates only.

The paper is organized as follows. In Section~\ref{framew} we review the overall calculational framework for CEP in high-energy $pp$ and $p \bar p$ collisions. Section~\ref{quark} focuses on heavy quarkonium production. After a careful treatment of the ingredients of the theoretical calculation, we present predictions for $\chi_{qJ}$ and $\eta_q$ production, with $q=c,b$ and $J=0,1,2$, at the Tevatron, RHIC and LHC. In Section~\ref{gamCEP} we update our previous study of $\gamma\gamma$ production, paying particular attention to formally subleading higher angular momentum states and the $\pi^0\pi^0$ background, and results are presented for Tevatron and LHC energies. Finally, Section~\ref{conc} contains a summary and outlook.

\section{Calculation framework and remarks}\label{framew}
\begin{figure}[b]
\begin{center}
\includegraphics[scale=0.8]{plots/kfig3.epsi}
\caption{The perturbative mechanism for the exclusive process $pp \to p+\chi+p$, with the eikonal and enhanced survival factors 
shown symbolically.}
\label{fig:pCp}
\end{center}
\end{figure} 
To calculate the CEP cross section for the general process $pp(\overline{p})\to p\;+\;X\;+\;p(\overline{p})$ we use the formalism of \cite{Khoze00a,Kaidalov03,Khoze00}. The amplitude is described by the diagram shown in Fig.~\ref{fig:pCp},
where the hard subprocess $gg \to X$ is initiated by gluon-gluon fusion and a second $t$-channel gluon is needed to screen the colour flow across the rapidity gap intervals. We can write the Born amplitude in the $Q_\perp$ factorised form~\cite{Khoze04,KKMRext}:
\begin{equation}\label{bt}
T=\pi^2 \int \frac{d^2 {\bf Q}_\perp\, \mathcal{M}}{{\bf Q}_\perp^2 ({\bf Q}_\perp-{\bf p}_{1_\perp})^2({\bf Q}_\perp+{\bf p}_{2_\perp})^2}\,f_g(x_1,x_1', Q_1^2,\mu^2;t_1)f_g(x_2,x_2',Q_2^2,\mu^2;t_2) \; ,
\end{equation}
where $\mathcal{M}$ is the colour-averaged, normalised sub-amplitude for the $gg \to X$ process:
\begin{equation}\label{Vnorm}
\mathcal{M}\equiv \frac{2}{M_X^2}\frac{1}{N_C^2-1}\sum_{a,b}\delta^{ab}q_{1_\perp}^\mu q_{2_\perp}^\nu V_{\mu\nu}^{ab} \; .
\end{equation}
Here $M_X$ is the central object mass, $a$ and $b$ are colour indices, $V_{\mu\nu}^{ab}$ is the $gg \to X$ vertex and $q_{i_\perp}$ are the transverse momenta of the incoming gluons, given by
\begin{equation}
q_{1_\perp}=Q_\perp-p_{1_\perp}\,, \qquad
q_{2_\perp}=-Q_\perp-p_{2_\perp}\,.
\label{qperpdef}
\end{equation}
The loop integral is cutoff for $|Q_\perp|,|q_{i_\perp}|<0.85$ GeV, as explained in~\cite{HarlandLang09,Khoze04}: clearly below (approximately) this scale we cannot trust perturbative QCD, and so to be conservative we cut off the integral to include only the perturbative contribution to the cross section, neglecting the contribution when any of the gluon propagators become too soft. The $f_g$'s in (\ref{bt}) are the skewed unintegrated gluon densities of the proton at the hard scale $\mu$, taken typically to be of the order of the produced massive state, i.e. $M_X/2$ in the examples which follow, and only one transverse momentum scale is taken into account by the prescription
\begin{align}\nonumber
Q_1 &= {\rm min} \{Q_\perp,|({\bf Q_\perp}-{\bf p}_{1_\perp})|\}\;,\\ \label{minpres}
Q_2 &= {\rm min} \{Q_\perp,|({\bf Q_\perp}+{\bf p}_{2_\perp})|\} \; .
\end{align}
The longitudinal momentum fractions carried by the gluons satisfy
\begin{equation}\label{xcomp}
\bigg(x' \sim \frac{Q_\perp}{\sqrt{s}}\bigg)  \ll \bigg(x \sim \frac{M_X}{\sqrt{s}}\bigg) \; .
\end{equation} 
The $t$ dependence of the $f_g$'s is not well known, but in the limit that the protons scatter at small angles, we can assume a factorization of the form
\begin{equation}\label{fnt}
f_g(x,x',Q^2,\mu^2;t)=f_g(x,x',Q^2,\mu^2)\,F_N(t) \; ,
\end{equation}
where the $t$-dependence is isolated in a proton form factor, which we take to have the phenomenological form $F_N(t)={\rm exp}(bt/2)$, with $b=4\,{\rm GeV}^{-2}$. 

\begin{figure}[h]
\begin{center}
\includegraphics[scale=0.7]{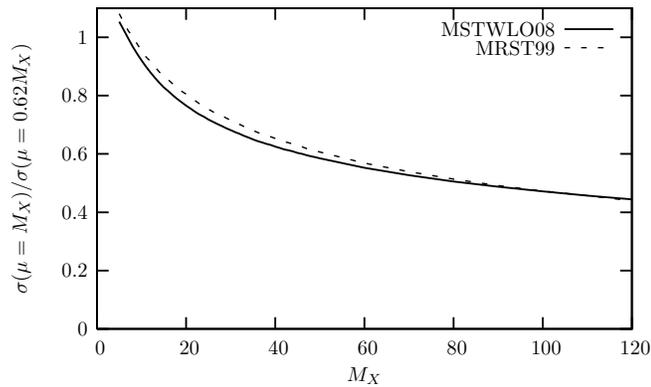}
\caption{Ratio of the cross sections for the CEP of a system of mass $M_X$, evaluated with $\Delta=k_\perp/(k_\perp+M_X)$ and $\Delta=k_\perp/(k_\perp+0.62M_X)$, using a modified form of Eq.~(26) of Ref.~\cite{Martin01ms}, as described in the text.}\label{mcomp}
\end{center}
\end{figure}

In the kinematic region specified by (\ref{xcomp}), the skewed unintegrated densities are given in terms of the conventional (integrated) densities $g(x,Q_i^2)$. To single log accuracy, we have\footnote{In actual calculations, we use a more precise phenomenological form given by Eq.~(26) of~\cite{Martin01ms}.}
\begin{equation}\label{fgskew}
f_g(x,x',Q^2,\mu^2)=R_g\frac{\partial}{\partial \, {\rm log} \, Q^2} \big[x g(x,Q^2)\sqrt{T_g(Q^2,\mu^2)}\big] \; ,
\end{equation}
where $T_g$ is the usual Sudakov factor which ensures that the active gluon does not emit additional real partons in the course of the evolution up to the hard scale $\mu$, so that the rapidity gaps survive. $R_g$ is the ratio of the skewed $x' \ll x$ unintegrated gluon distribution to the conventional diagonal density $g(x,Q^2)$. For $x \ll 1$ it is completely determined~\cite{Shuvaev99}. The explicit form for $T_g$ is given by resumming the virtual contributions to the DGLAP equation. It is given by
\begin{equation}\label{ts}
T_g(Q_\perp^2,\mu^2)={\rm exp} \bigg(-\int_{Q_\perp^2}^{\mu^2} \frac{d {\bf k}_\perp^2}{{\bf k}_\perp^2}\frac{\alpha_s(k_\perp^2)}{2\pi} \int_{0}^{1-\Delta} \bigg[ z P_{gg}(z) + \sum_{q} P_{qg}(z) \bigg]dz \bigg) \; .
\end{equation}
Here we fix the upper cutoff on the $z$ integral in order to correctly resum the single soft ${\rm ln}(1-z)$ terms due to wide angle soft gluon emission. It was previously claimed in Ref.~\cite{Kaidalov03} that this could be achieved by taking the upper limit to be
\begin{equation}\label{deltao}
\Delta_{\rm old} =\frac{k_\perp}{k_\perp+0.62 M_X} \; .
\end{equation}
However, as has been shown in \cite{Coughlin09}, this result is incorrect, and the correct choice of cutoff is in fact
\begin{equation}\label{deltan}
\Delta=\frac{k_\perp}{k_\perp+M_X} \; .
\end{equation}
i.e. of the same form as (\ref{deltao}), but with the replacement $0.62 M_X \to M_X$. As has been noted in~\cite{Coughlin09}, the correct inclusion of these single logarithms is quite significant, giving a factor of $\sim 2$ decrease in the cross section for the CEP of a $M=120\;{\rm GeV}$ Higgs boson at $\sqrt{s}=14 \;{\rm TeV}$. Note that this replacement changes {\it both} $\sqrt{T_g}$ and the derivative $\partial T_g/\partial\log Q^2$ in (\ref{fgskew}). A smaller  $\Delta$ in (\ref{deltan}) means a larger rapidity (angular) interval is allowed for soft gluon emission, which leads to a smaller value of $T_g$ (i.e. a smaller probability not to emit an additional gluon, that is a stronger negative virtual loop correction). However simultaneously the probability for {\it real} emission, after which the active gluon gets transverse momentum $Q_\perp$, increases. The last effect is described by the positive $\partial T_g/\partial\log Q^2$ term which grows as $\Delta$ decreases. Thus the replacement $0.62M_X\to M_X$ will affect the expected CEP cross section in a way that depends non-trivially on the central object mass $M_X$ as well as the $x$ ($\sqrt{s}$) values being considered. We show the $M_X$ dependence explicitly in Fig.~\ref{mcomp}, where we plot the ratio of the cross sections with $\Delta$ defined as in (\ref{deltao}) and (\ref{deltan}) as a function of $M_X$, and using the MRST99~\cite{Martin99} and MSTW08LO~\cite{Martin09} PDF sets (from which we can see that the PDF dependence of the ratio is negligible). Clearly the effect becomes less severe as we go to lower masses: indeed, the change in the $\chi_c$ rate, as calculated in~\cite{HarlandLang09,Khoze04}, is only of order $\lesim 10\%$ and may lead to a small cross section increase. Such an effect is clearly well within the other (large) theoretical uncertainties associated with the CEP of low $M_X$ objects, but the important point is that for the lower $M_X$ standard candle processes we will be considering in this paper, the effect from the replacement of (\ref{deltan}) is numerically not too large.\footnote{Note also that the changes caused by the modified cutoff $\Delta$ are at least partly compensated by other effects, such as the NLO corrections to the unintegrated gluon density (see \cite{MRW}) and accounting for the self-energy insertions in the
propagator of the screening gluon.}

For completeness we note that in practice we replace the expression for $f_g$ in (\ref{fgskew}) by the more accurate phenomenological fit of~\cite{Martin01ms}. Since this was formulated for the case $\Delta=k_\perp/(k_\perp+M_\chi/2)$, we must take a little care to correctly include the replacement of (\ref{deltan}). In particular, the new $\Delta$ is not only included implicitly in (26) of Ref.~\cite{Martin01ms} via its effect on the $\sqrt{T_g}$ term, but also explicitly by making the replacement $\mu \to M_X$ in the ${\rm ln}((\mu+k_t/2)/k_t)$ term, so as to correctly include the leading logarithmic contribution coming from
\begin{equation}\label{LL}
\frac{\partial \sqrt{T_g}}{\partial {\rm ln}\,Q^2}\stackrel{(LL)}{=}\frac{N_C \alpha_S}{2\pi}{\rm ln}\bigg(\frac{1}{\Delta}\bigg)
\end{equation}
in (\ref{fgskew}). Such a procedure was also used in~\cite{HarlandLang09} for the $\Delta$ value of (\ref{deltao}). While this prescription does not exactly reproduce the ratio $\sigma(\Delta(0.62 M_X))/\sigma(\Delta(M_X))$ found using (\ref{fgskew}), for which the replacement of (\ref{deltan}) will also have a subleading effect on the non-logarithmic contributions to (\ref{LL}), it agrees approximately (i.e. to within the formal accuracy of (\ref{fgskew})), with the agreement improving as we go to the lower masses $M_X$ that we will be considering in this study.
\section{Central exclusive heavy quarkonium production}\label{quark}
\subsection{Vertex calculation}\label{vertex}

The calculation of the CEP cross sections for $\chi_{c0}$, $\chi_{c1}$ and $\chi_{c2}$ production is explained in detail in~\cite{HarlandLang09}, and we only review the important aspects here. The cross section is given by (\ref{bt}), where $\mathcal{M}$ ($\equiv V_J$, in the previous notation) depends on the spin $J$ and parity $P$ of the centrally produced particle, and is readily calculated by a simple extension of the formalism of~\cite{Kuhn79}, where the coupling of $P$-wave quarkonium states to two off-mass-shell photons is considered. The relevant calculation for $\chi_b$ production then proceeds in exact analogy to the $\chi_c$ case, the only difference being the input masses $M_\chi$ and widths $\Gamma(\chi \to gg)$. We now also consider pseudoscalar $\eta_{c,b}$ production, which can be calculated using the same formalism as for $\chi_c$ production. The $gg\to\chi,\eta$ vertices are given by~\cite{HarlandLang09}
\begin{align}\label{V0}
&V_{0^+}=\sqrt{\frac{1}{6}}\frac{c_\chi}{M_\chi}(3M_\chi^2(q_{1_{\perp}}q_{2_{\perp}})-(q_{1_{\perp}}q_{2_{\perp}})(
q_{1_{\perp}}^2+q_{2_{\perp}}^2)-2q_{1_{\perp}}^2q_{2_{\perp}}^2)  \; ,\\ \label{V1}
&V_{1^+}=-\frac{2ic_\chi}{s} p_{1,\nu}p_{2,\alpha}((q_{2_\perp})_\mu(q_{1_\perp})^2-(q_{1_\perp})_\mu(q_{2_\perp})^2)\epsilon^{\mu\nu\alpha\beta}\epsilon^{*\chi}_\beta  \; ,\\
\label{V2}
&V_{2^+}=\frac{\sqrt{2}c_\chi M_\chi}{s}(s(q_{1_\perp})_\mu(q_{2_\perp})_\alpha+2(q_{1_\perp}q_{2_\perp})p_{1\mu}p_{2\alpha})\epsilon_\chi^{*\mu\alpha}  \; ,\\ \label{V0m}
&V_{0^-}=ic_\eta(q_{1_\perp}\times q_{2_\perp})\cdot n_0\;,
\end{align}
where $q_{i_\perp}$ are the incoming gluon momenta, given by (\ref{qperpdef}), and $n_0$ is a unit vector in the direction of the colliding hadrons (in the c.m.s frame). The normalisation factors are given by
\begin{equation}
c_\chi=\frac{1}{2\sqrt{N_C}}\frac{16 \pi \alpha_S}{(q_1 q_2)^2}\sqrt{\frac{6}{4\pi M_\chi}}\phi'_P(0), \qquad
c_\eta=\frac{1}{\sqrt{N_C}}\frac{4 \pi \alpha_S}{(q_1 q_2)}\frac{1}{\sqrt{\pi M_\eta}}\phi_S(0)\;,
\end{equation}
where $\phi_{S(P)}(0)$ is the $S(P)$-wave wavefunction at the origin. Clearly the $\eta$ vertex $V_{0^-}$ vanishes in the limit of $p_\perp=0$, as we expect from the $J^P_z=0^+$ selection rule~\cite{Khoze00a,Kaidalov03}. Moreover, at small $p_\perp$ we will have
\begin{equation}
|V_{0^-}|^2 \sim p_{1_\perp}^2 p_{2_\perp}^2 \sin^2\phi\;,
\end{equation}
where $\phi$ is the azimuthal angle between the outgoing protons. The $\eta$ CEP cross section will therefore be heavily suppressed relative to the $\chi_0$ rate by a factor of $\sim \langle \mathbf{p}_{\perp}^2\rangle^2/2\langle \mathbf{Q}_{\perp}^2\rangle^2$, i.e. roughly two orders of magnitude. However, as the $\chi$ cross section depends on the value of $\phi'_P(0)$ while the $\eta$ cross section depends on $\phi_S(0)$, this can only be considered as a very rough estimate. Indeed, if we normalise $|V_{0-}|^2$ in terms of the $\eta_c \to gg$ width and $\eta_c$ mass, as in Eq.~(\ref{etanorm}) of Section~\ref{norm}, then the higher experimental value of the $\eta_c$ width and the lower $\eta_c$ mass will partly compensate (by a factor of $\sim 4$) this suppression.

For $\chi_{1,2}$ production, the level of expected suppression relative to the $\chi_0$ rate can be estimated by~\cite{HarlandLang09}
\begin{equation}\label{comp}
|V_{0^+}|^2:|V_{1^+}|^2:|V_{2^+}|^2 \sim 1:\frac{\left\langle \mathbf{p}_{\perp}^2\right\rangle}{M_\chi^2}:\frac{\left\langle \mathbf{p}_{\perp}^2\right\rangle^2}{\left\langle \mathbf{Q}_\perp^2\right\rangle^2}\; .
\end{equation}
If for simplicity we assume $\mathbf{Q}_\perp^2 \approx 1.5\, {\rm GeV}^2$, $M_{\chi_c}^2\approx 10\, {\rm GeV}^2$ and  $M_{\chi_b}^2\approx 100\, {\rm GeV}^2$ we then obtain
\begin{align}\label{roughc}
|V_{0^+}|^2:|V_{1^+}|^2:|V_{2^+}|^2 &\sim 1:\frac{1}{40}:\frac{1}{36}\quad (c\overline{c}) \;,\\ \label{roughb}
|V_{0^+}|^2:|V_{1^+}|^2:|V_{2^+}|^2 &\sim 1:\frac{1}{400}:\frac{1}{36}\quad (b\overline{b}) \; .
\end{align}
Note that these are only rough estimates which should be confirmed by explicit calculation -- see below. Most significantly, we can see that we would expect $\chi_{b1}$ states to give a negligible contribution to the overall $\chi_b$ CEP rate. In fact, for the higher mass $b\overline{b}$ states we expect a slightly higher $\left\langle \mathbf{Q}_\perp^2\right\rangle$ value and so a slightly stronger level of suppression for $\chi_{b2}$ production than in the $c\overline{c}$ case. However, as with $\chi_c$ CEP, there remains the possibility that $\chi_{b2}$ states may contribute to $\chi_b$ production via the $\chi_b \to \Upsilon\gamma$ decay chain, although the precise value of the $0^+/2^+$ ratio depends not only on the usual uncertainties of the calculation but also on the value taken for the $\chi_{b0} \to \Upsilon\gamma$ branching ratio (see Section \ref{norm} for a discussion of this).

\subsection{Normalisation and uncertainties}\label{norm}

As in the case of $\chi_c$ production we can normalise the $\eta_c$ vertex to the $\eta_c \to gg$ width, with (in the $q_{i_\perp}^2\ll M_\eta^2$ limit)
\begin{equation}\label{etanorm}
|V_{0^-}|^2=\frac{8\pi\Gamma(\eta \to gg)}{M_\eta^3}\, \cdot\,|{\bf q}_{1_\perp}\times {\bf q}_{2_\perp}|^2\;,
\end{equation}
and the value for the width taken from data by assuming $\Gamma(\eta_c \to gg) \approx \Gamma_{\rm tot}(\eta_c)=27.4$ MeV~\cite{PDG} (this assumes the same K--factor for the $\eta \to gg$ and $gg \to \eta$ vertices, which is only true to a certain degree of approximation). In the $\eta_b$ case, there is no experimental value for the total decay width, but we can nevertheless use data to normalise the $gg \to \eta_b$ vertex by noting that, in the `static quark' limit~\cite{Barbieri75,Bergstrom79},
\begin{equation}
\Gamma(\eta_b \to gg) = \frac{2}{9e_Q^4}\bigg(\frac{\alpha_S}{\alpha}\bigg)^2\Gamma(\eta_b \to \gamma\gamma)=\frac{2}{3e_Q^2}\bigg(\frac{\alpha_S}{\alpha}\bigg)^2\bigg(\frac{M_\Upsilon}{M_\eta}\bigg)^2\Gamma(\Upsilon \to \mu^+\mu^-)\;,
\end{equation}
where $e_Q=1/3$ and we have assumed the same value of the wavefunction at the origin for the $L=0$ pseudoscalar and vector states. The $\mathcal{O}(\alpha_S)$ QCD corrections to these relations are known~\cite{Fabiano02,Parmar10}, and from experiment we have $\Gamma(\Upsilon \to \mu^+\mu^-)=1.34$ keV~\cite{PDG}.

When considering $\chi_b$ production we must unfortunately deal with other quite sizeable uncertainties that were not present in the $\chi_c$ case. First, to normalise the $gg \to \chi_b$ vertex we require a value for the $\chi_b \to gg$ width. For $\chi_c$ production, one way to achieve this is to assume $\Gamma(\chi_{c0} \to gg)\approx\Gamma_{\rm tot}(\chi_c)$, which we can then take from data~\cite{PDG}. However no such data exists for the $\chi_b$ and so we must take a value from the theoretical literature, such as potential models or lattice calculations. If we consider the available potential model estimates (see for example~\cite{Kwong88,Laverty09,Eichten95}), there exists quite a wide spread in the predicted values, with $\Gamma(\chi_{b0} \to gg)$ (including the usual NLO K--factor) ranging over roughly $0.8-3$~MeV. We will take $\Gamma(\chi_{b0} \to gg)=0.8$~MeV from~\cite{Kwong88} as our benchmark value, as it is consistent with both lattice estimates (\cite{Kim95} gives the value $K\times 0.35$~MeV $\lesim 0.8$~MeV), and the fact that our calculation of the Lorentz structure of the $gg \to \chi$ vertices (Eqs.~(\ref{V0}--\ref{V2})) depends on the non-relativistic potential model approach that is used in Ref.~\cite{Kwong88}. Nevertheless the wide range in predictions for $\Gamma(\chi_b \to gg)$ clearly represents a sizeable uncertainty in the normalisation of the $\chi_b$ CEP cross section.

A further uncertainty that we must deal with when considering the $\chi_{b0} \to \Upsilon\gamma (\to \mu^+\mu^-\gamma)$ decay chain is that there exists no experimental determination of the $\chi_{b0} \to \Upsilon\gamma$ branching ratio: only an old value for the upper bound of ${\rm BR}(\chi_{b0} \to \Upsilon\gamma)<6\%$ is known, determined by Crystal Ball collaboration in 1986 \cite{cball}. Given that we expect the $\chi_{b0}$ state to dominate the $\chi_b$ CEP process, this will represent a large uncertainty in the total predicted $\chi_b$ cross section via the $\chi_{b0} \to \Upsilon\gamma \to \mu^+\mu^-\gamma$ channel, although using the upper bound on the branching ratio we can of course predict an approximate upper bound on the $\chi_b$ CEP cross section via this decay chain. We must again take a specific value from the theoretical literature to use as an input parameter for our calculation, and for overall consistency we again take the prediction of~\cite{Kwong88} of ${\rm BR}(\chi_{b0} \to \Upsilon\gamma)\approx 3\%$. The predictions of Ref.~\cite{Kwong88} for the $\chi_{b1,2} \to \Upsilon\gamma$ branching ratios are in very good agreement with the data and indeed, to the authors' knowledge, few other predictions for the $\chi_{b0}$ branching ratio exist. We can then compare this with the experimentally determined branching ratios~\cite{PDG}
\begin{align}
{\rm BR}(\chi_{b1} \to \Upsilon\gamma)&=(35 \pm 8)\%\;,\\
{\rm BR}(\chi_{b2} \to \Upsilon\gamma)&=(22 \pm 4)\%\;.
\end{align}
Again we can see the same strong hierarchy in branching ratios as with the $\chi_c \to J/\psi \gamma$ decays, which may compensate the initial suppression of (\ref{roughb}) for $\chi_{b2}$ production. 

While the level of uncertainty in $\Gamma(\chi_b \to gg)$ and ${\rm BR}(\chi_{b0} \to \Upsilon\gamma)$ is significant, an important point is that to a good degree of accuracy the predicted cross section for $\chi_b$ CEP via the $\chi_b \to \Upsilon\gamma$ decay chain does not depend on the $\chi_b$ total widths, as the dependence cancels once the $\chi_b$ cross section is multiplied by the $\chi_b \to \Upsilon\gamma$ branching ratio. As well as the uncertainty in the NLO corrections to the $gg \to \chi$ vertex, which we discuss below, there remains the uncertainty in the $\Gamma(\chi_b \to \Upsilon\gamma)$ widths, although fairly consistent theoretical predictions for these exist (the potential model calculations of Refs.~\cite{Kwong88,Radford07} give results that are in approximate agreement, for example).

A final important source of uncertainty, previously discussed in Ref.~\cite{HarlandLang09}, is in the NLO K--factors for the $\chi/\eta \to gg$ widths. First, we have assumed that the corrections to the $\chi/\eta \to gg$ widths and the $gg \to \chi/\eta$ widths are the same but, as is well known, this is not exactly true and so the K--factors we use can only be considered as rough estimates of the expected NLO corrections. By taking $\Gamma(\chi_0 \to gg) \approx \Gamma_{\rm tot}(\chi_0)$ to normalise the $gg \to \chi_0$ vertex we implicitly include such a K-factor, but the expected NLO corrections will in general depend on the spin of the produced state, and so we should be careful in using this value to normalise the $\chi_{1,2} \to gg$ vertices. In particular, as discussed in footnote 10 of~\cite{HarlandLang09}, we have reason to expect the K-factor for the $gg \to \chi_1$ vertex to be close to unity, while in the $\chi_{(c,b)2}$ case the calculated K--factors are considerably smaller than the corresponding $\chi_{(c,b)0}$ values (see~\cite{Parmar10} and references therein). However, we note that these K--factors are calculated for a spin-averaged initial state $\chi_2$, while for CEP the different $\chi_2$ helicities receive different weights (in particular, the $|J_Z|=2$ state is dominantly produced), and so we cannot reasonably assume the same correction for the $\chi_2 \to gg$ and $gg \to \chi_2$ widths. For simplicity we will therefore assume $K_{\rm NLO}(\chi_0)\approx 1.5$ and $K_{\rm NLO}(\chi_{1,2})\approx 1$ for the $\chi_{c,b}$ states. However, such an assumption can only be loosely justified, and so these spin-dependent NLO corrections represent an important source of uncertainty, in particular in the predictions for the relative $\chi_0$ to $\chi_{1,2}$ cross sections. Finally, we note that it is well known (see for example \cite{Barbieri:1980yp}) that the NNLO and higher-order radiative corrections to the $\chi \to gg$ transition could be numerically large, which may result in further uncertainties in the theoretical predictions.

\subsection{Particle distributions and inclusion of soft survival effects}\label{surveff}
In~\cite{HarlandLang09}, the effect of allowing a non-zero proton $p_\perp$ in the cross section calculation, which we note is essential for the case of $\chi_{1,2}$ and $\eta$ CEP, was modelled by fitting the slope $b$ of the proton form factor to the $p_{\perp_X}$ distribution of the centrally production object $X$, which could then be used to estimate the expected eikonal survival probability $\langle S^2_{\rm eik}\rangle$, averaged over the $p_\perp$ of the outgoing protons, as well as allowing a simple Monte Carlo implementation. We now go beyond this approximation by explicitly performing the phase space integration over the $p_\perp$ dependent survival factor and subprocess cross section via
\begin{equation}\label{ampnew}
\frac{{\rm d}\sigma}{{\rm d} y_X}=\int {\rm d}^2\mathbf{p}_{1_\perp} {\rm d}^2\mathbf{p}_{2_\perp} \frac{|T(\mathbf{p}_{1_\perp},\mathbf{p}_{2_\perp}))|^2}{16^2 \pi^5} S_{\rm eik}^2(\mathbf{p}_{1_\perp},\mathbf{p}_{2_\perp})\; ,
\end{equation}
where $T$ is given by (\ref{bt}) and the gap survival factor is most simply written in impact parameter $b_t$ space as
\begin{equation}\label{seik}
S^2_{\rm eik}(b_t)=\exp(-\Omega(s,b_t))\;.
\end{equation}
Here $b_t$ is the separation in the transverse plane between the centres of the colliding protons and $\Omega(s,b_t)$ is the proton opacity, which is the Fourier transform of the two-particle ($s$--channel) {\it irreducible} amplitude $A(s,q_t)$: physically, $\exp(-\Omega(s,b_t))$ represents the probability that no inelastic scattering occurs at impact parameter $b_t$. In the `single channel eikonal' model of soft rescattering, which we will for simplicity consider here, it is related to the elastic amplitude by
\begin{equation}
T_{\rm el}(s,b_t)=i\big(1-e^{-\Omega(s,b_t)/2}\big)\;.
\end{equation}
where we neglect for simplicity the imaginary part of $\Omega(s,b_t)$ (at high energies ${\rm Re}\,T_{\rm el}/{\rm Im}\,T_{\rm el}$ is small and can be evaluated via a dispersion relation in the complete treatment). Explicitly, working in momentum space we must calculate the CEP amplitude including rescattering effects $T^{\rm res.}$ by integrating over the transverse momentum ${\bf k}_\perp$ carried round the Pomeron loop (represented by `$S_{\rm eik}^2$' in Fig.~\ref{fig:pCp}). The ${\bf k}_\perp$ dependence of the screening amplitude is given by the Fourier transform of $T_{\rm el}$, as given in terms of the proton opacity $\Omega(s,b_t)$, while the ${\bf p}_\perp$ dependence of the `bare' amplitude $T(\mathbf{p}_{1_\perp},\mathbf{p}_{2_\perp})$ is calculated explicitly within the perturbative model. The amplitude including rescattering corrections is then given by
\begin{equation}
T^{\rm res}(\mathbf{p}_{1_\perp},\mathbf{p}_{2_\perp}) = \frac{i}{s} \int\frac{{\rm d}^2 \mathbf {k}_\perp}{8\pi^2} \;T_{\rm el}({\bf k}_\perp) \;T(\mathbf{p'}_{1_\perp},\mathbf{p'}_{2_\perp})\;,
\end{equation}
where $\mathbf{p'}_{1_\perp}=({\bf k}_\perp-{\bf p}_{1_\perp})$ and $\mathbf{p'}_{2_\perp}=({\bf k}_\perp+{\bf p}_{2_\perp})$. We must then add this to the `bare' amplitude excluding rescattering effects to give the full physical amplitude, which we can square to give the CEP cross section including eikonal survival effects
\begin{equation}
\frac{{\rm d}\sigma}{{\rm d} y_X} \propto \int {\rm d}^2\mathbf{p}_{1_\perp} {\rm d}^2\mathbf{p}_{2_\perp} |T(\mathbf{p}_{1_\perp},\mathbf{p}_{2_\perp})+T^{\rm res}(\mathbf{p}_{1_\perp},\mathbf{p}_{2_\perp})|^2 \;.
\end{equation}
In general there is clearly a non-trivial interference between the bare and screened amplitude that will depend on the choice of soft rescattering model as well as the particular hard process $gg\to X$. Finally, to make contact with the notation of (\ref{ampnew}), we note that
\begin{equation}
S_{\rm eik}^2(\mathbf{p}_{1_\perp},\mathbf{p}_{2_\perp})\equiv \frac{|T(\mathbf{p}_{1_\perp},\mathbf{p}_{2_\perp})+T^{\rm res}(\mathbf{p}_{1_\perp},\mathbf{p}_{2_\perp})|^2}{|T(\mathbf{p}_{1_\perp},\mathbf{p}_{2_\perp})|^2}\;.
\end{equation}
For further details and an explanation of the generalisation to the `two-channel eikonal' model that we use for our calculation we refer the reader to~\cite{KMRtag} and references therein. 

Explicitly including the ${\bf p}_\perp$ dependent soft survival effects in this way not only provides a more accurate prediction for the expected soft suppression, and hence the total CEP cross section, but also allows us to more precisely predict the final state particle distributions that we would expect to see in the detector, i.e. including the full effect of \textit{both} the subprocess cross section $\hat{\sigma}(gg \to X)$ and secondary rescatterings on the distributions.  This is clearly important if we are to confront these aspects of the model with existing and future data, in particular in the presence of the proposed forward proton detectors~\cite{AR, FP420, RP}. More specifically, we recall that a detailed study of the $p_\perp$ and azimuthal distributions of the outgoing protons would allow us to probe the different available models for soft diffraction, see for instance~\cite{KMRtag}.

\begin{figure}[t]
\begin{center}
\includegraphics[scale=0.5]{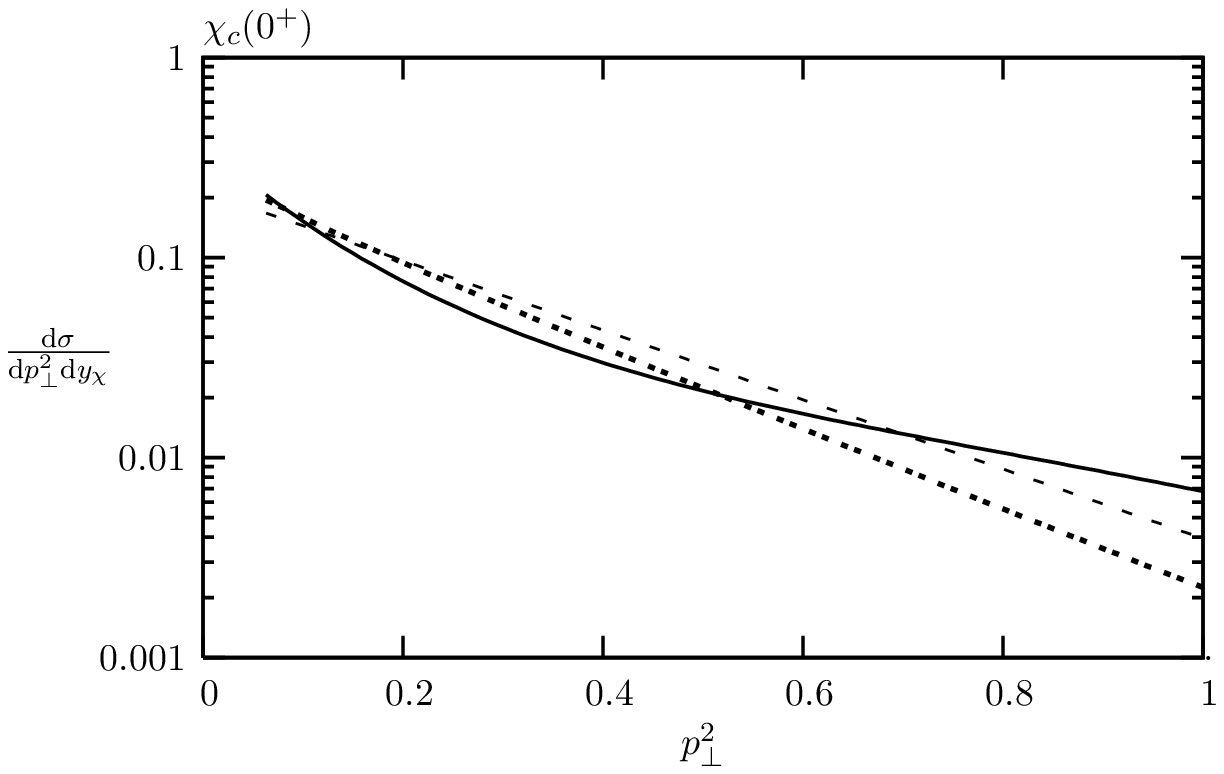}
\includegraphics[scale=0.5]{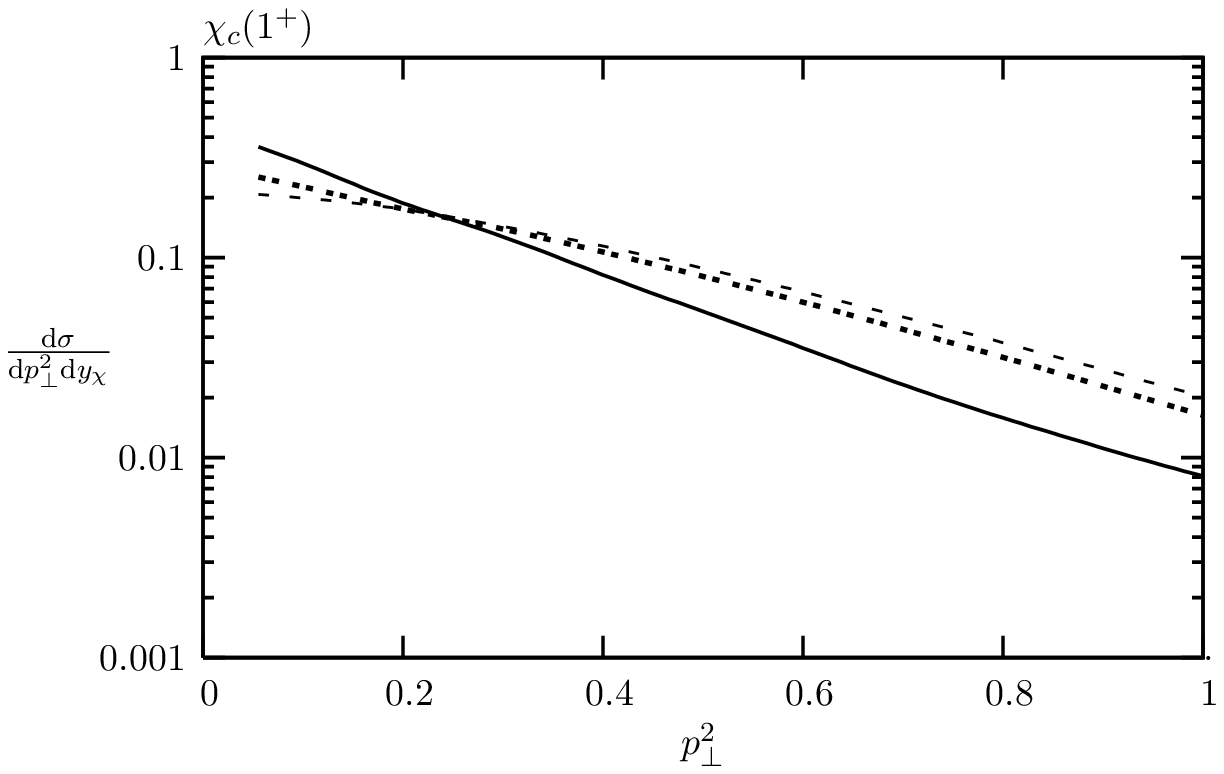}
\includegraphics[scale=0.5]{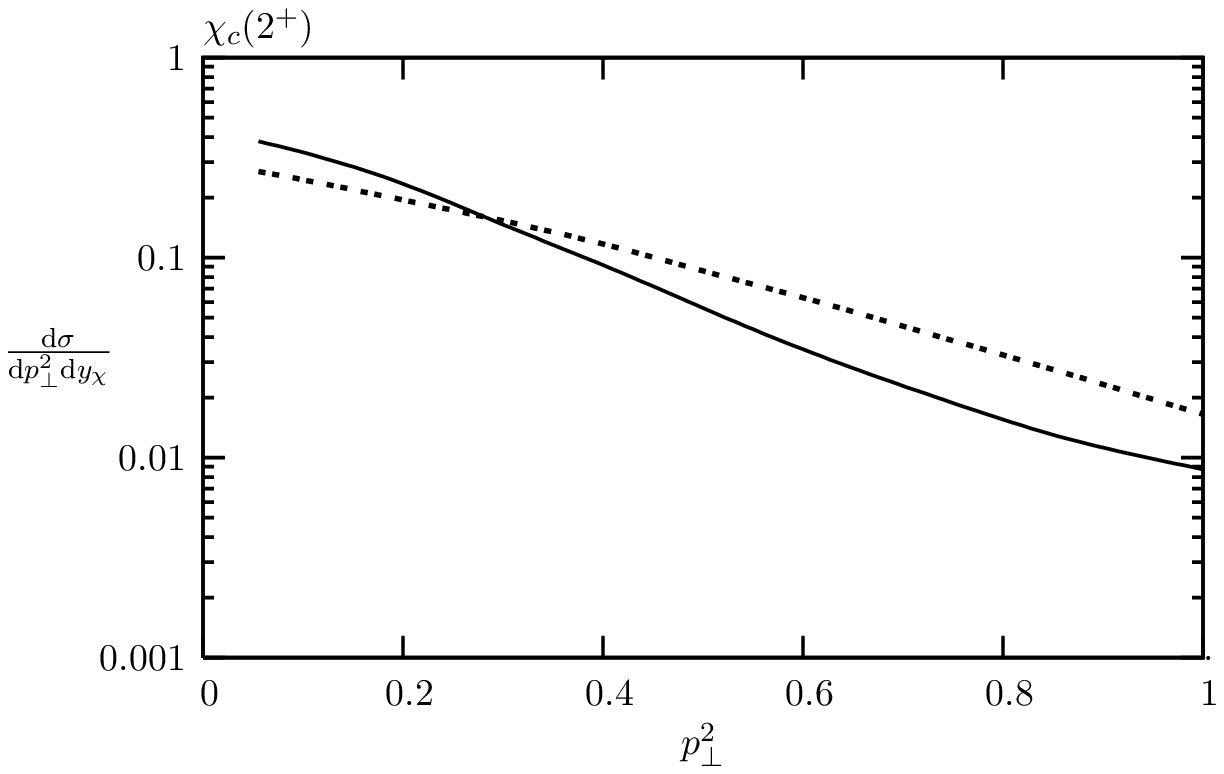}
\includegraphics[scale=0.5]{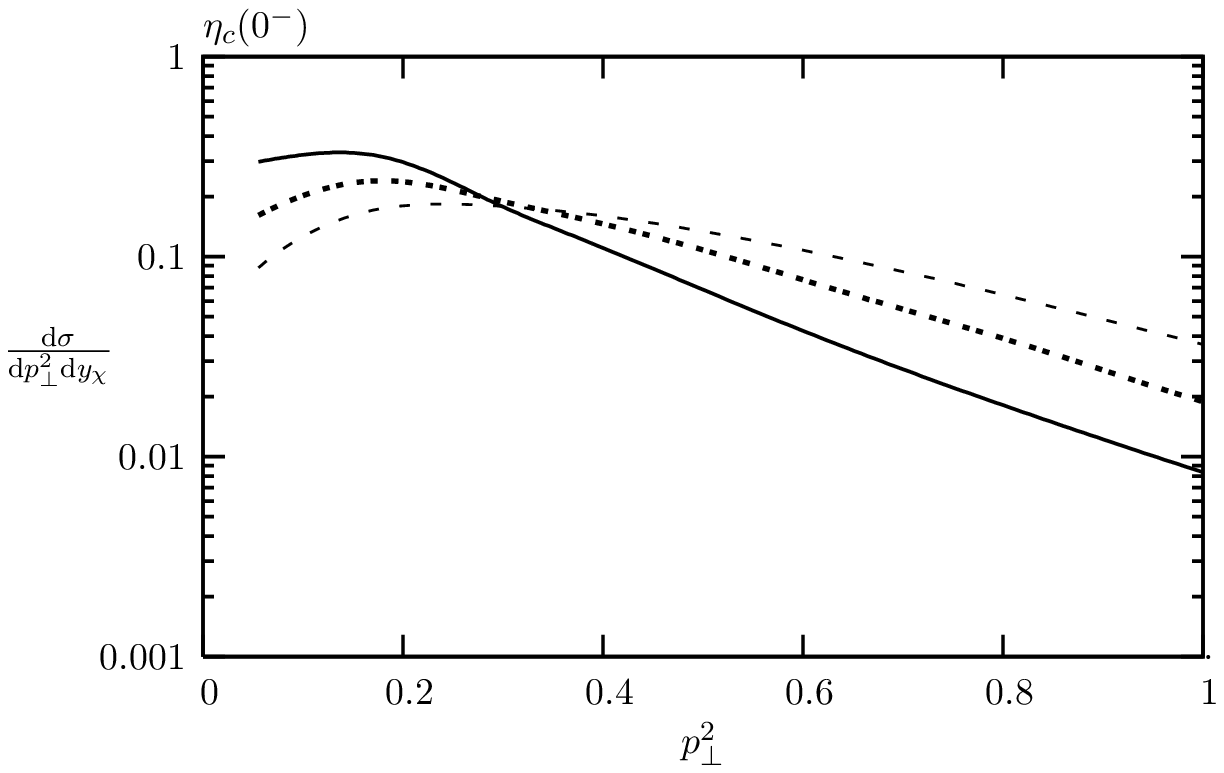}
\caption{Distribution (in arbitrary units) within the perturbative framework of the outgoing proton ${\bf p}_{1_\perp}^2$, integrated over the second proton ${\bf p}_{2_\perp}$ and at rapidity $y_X=0$, for the CEP of different $J^P$ $c\overline{c}$ states at $\sqrt{s}=14$ TeV. The solid (dotted) line shows the distribution including (excluding) the survival factor, calculated using the two channel eikonal model of Ref.~\cite{KMRsoft}, while the dashed line shows the distribution in the small $p_\perp$ limit, using the vertices of Eqs.~(\ref{R0p}--\ref{R0m}) and excluding the survival factor.}\label{surv2}
\end{center}
\end{figure}

We recall that the $p_\perp$ distribution of the outgoing protons is not just described by the exponential form factor $F_N(t)=\exp(bt)$ in (\ref{fnt}), with slope $b=4$ GeV$^{-2}$. First, an additional $p_\perp$ dependence is introduced in (\ref{bt}) when we account for the proton momenta $p_{1,2_\perp}$ in the gluon propagators and the matrix element $\mathcal{M}$ of the $gg \to X$ vertex. In \cite{HarlandLang09} it was shown that this inclusion of a non-zero proton $p_\perp$ in the $Q_\perp$ integral of (\ref{bt}) in general leads to a decrease in $\langle p_\perp^2 \rangle$, which we modelled by introducing an increased `effective' slope parameter $b_{\rm eff}$ with $b_{\rm eff}>b$. In particular, in the limit that ${\bf p}_\perp^2\ll {\bf Q}_\perp^2$, we expect (see~\cite{HarlandLang09,Kaidalov03}) the $gg \to \chi/\eta$ vertices to have the form
\begin{align}\label{R0p}
|V_{0^+}|^2 &\sim {\rm const.}\;,\\ \label{R1p} 
|V_{1^+}|^2 &\sim ({\bf p}_{1_\perp}-{\bf p}_{2_\perp})^2\;,\\ \label{R0m}
|V_{0^-}|^2 &\sim {\bf p}_{1_\perp}^2{\bf p}_{2_\perp}^2\sin^2{\phi}\;,
\end{align}
while there does not exist a simple closed form for the $\chi_2$ case, as detailed in Appendix \ref{distap}, where we also justify the above expressions. However we expect these to be changed by the $Q_\perp$ loop integral and non-zero $p_\perp$ effects, with the size of the correction being roughly of order $\sim p_\perp^2/\langle Q_\perp^2 \rangle$, and it is these corrections that are included in the modified $b_{\rm eff}$. However, this characterises the proton distribution corresponding to the {\it bare} CEP amplitude, that is without the inclusion of screening corrections. The slope $b_{\rm eff}$ will therefore still not describe the proton distributions we expect to be measured experimentally, which will be further modified by the gap survival factor $S_{\rm eik}^2$ as in (\ref{ampnew}). To estimate the role of this effect we may calculate the ratio 
\begin{equation}
 R_{\rm scr}=\frac{S_{\rm eik}^2(p_{1\perp}=p_{2\perp}=0)}{\langle S_{\rm eik}^2\rangle}\ .
\label{rscr}
\end{equation}
Since this accounts for the $p_\perp$ distributions of both protons, the experimentally observed slope will be given by $b_{\rm exp}\simeq \sqrt{R_{\rm scr}}\,b_{\rm eff}$. As in impact parameter space small transverse momenta $p_{1,2_\perp}$ correspond to large $b_t$, where the gap survival probability is higher, we will have $R_{\rm scr} >1$, and so we will typically expect the hierarchy $b_{\rm exp}>b_{\rm eff}>b$. That is, the inclusion of screening corrections induces a further steepening in the proton $p_\perp$ distribution. This hierarchy is clear in Fig.~\ref{surv2}, where we plot the $p_\perp$ distribution of the outgoing protons for $\chi_{c(0,1,2)}$ and $\eta_c$ production at the LHC ($\sqrt{s}=14$ TeV). The important point is that in all cases the curves calculated within the perturbative framework have steeper $p_\perp$ behaviour (at small $p_\perp$) than those calculated using Eqs.~(\ref{R0p}--\ref{R0m}) assuming the usual exponential proton form factor $\exp(bt)$, and this steepening is further increased upon the inclusion of screening effects. This also remains true for $\chi_{b(0,1,2)}$ and $\eta_b$ production and different c.m.s. energies.

\begin{figure}[t]
\begin{center}
\includegraphics[scale=0.5]{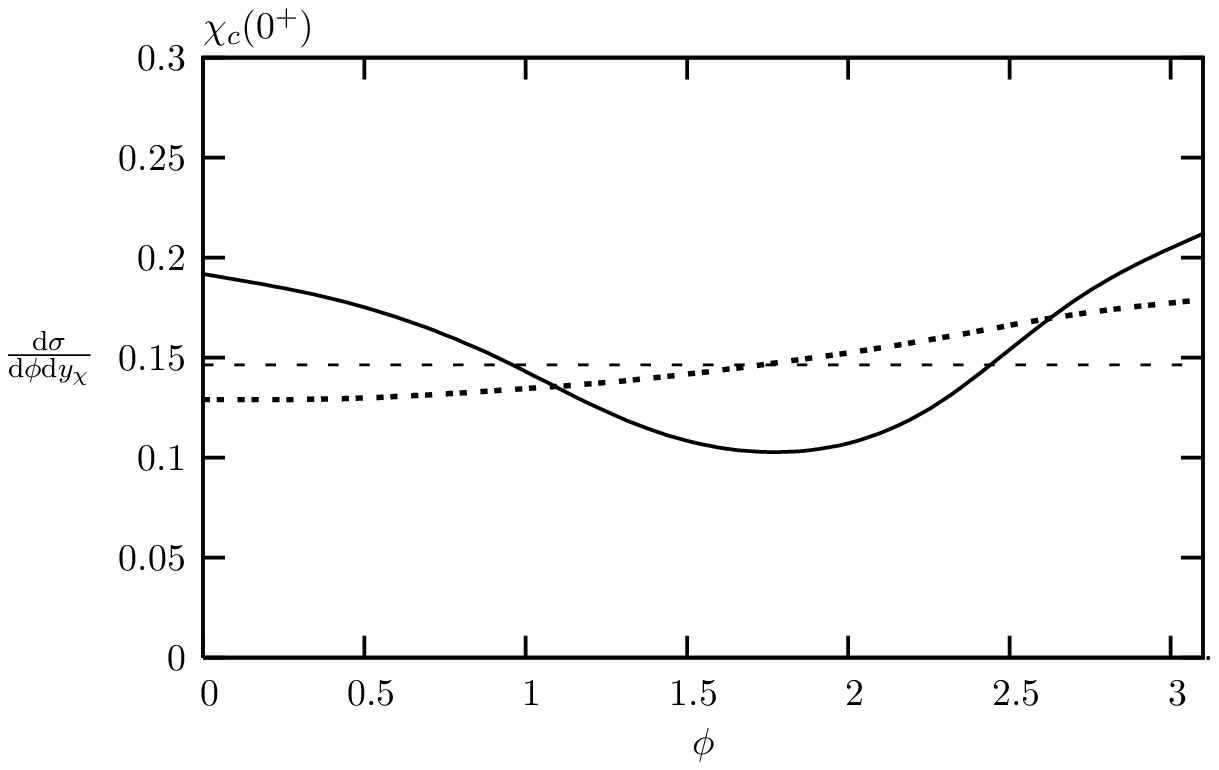}
\includegraphics[scale=0.5]{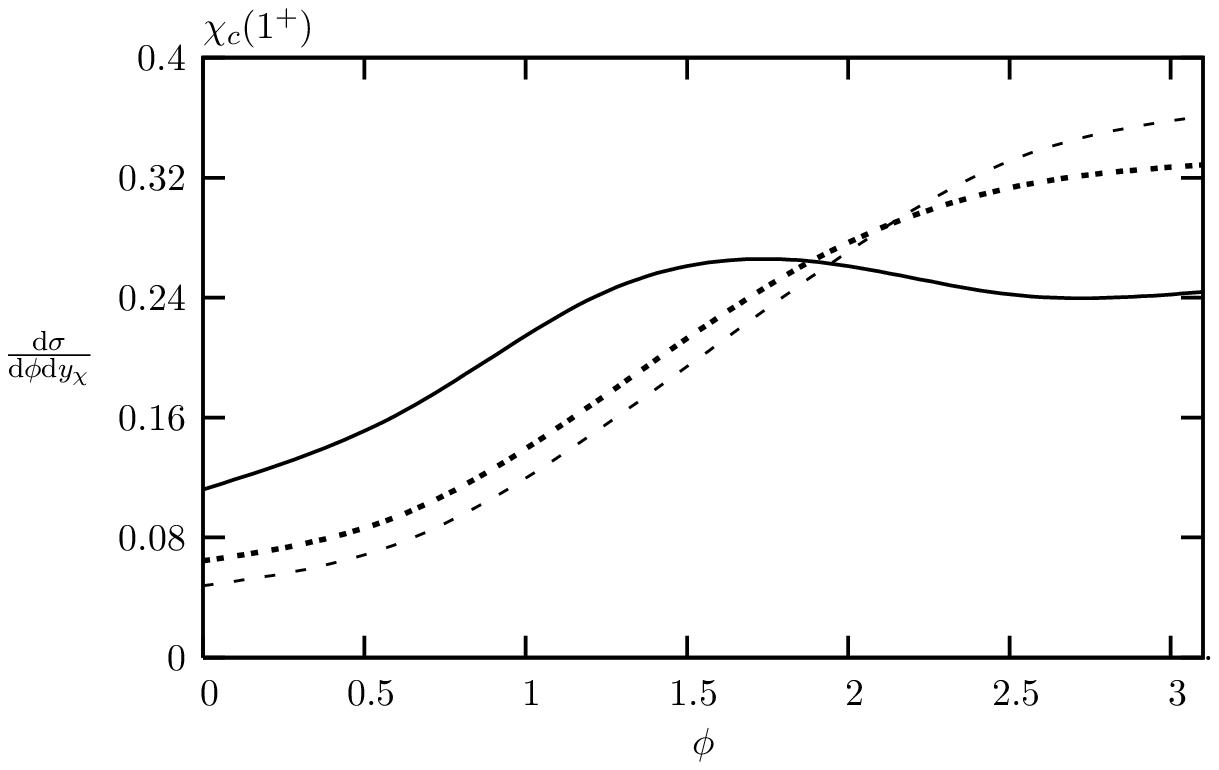}
\includegraphics[scale=0.5]{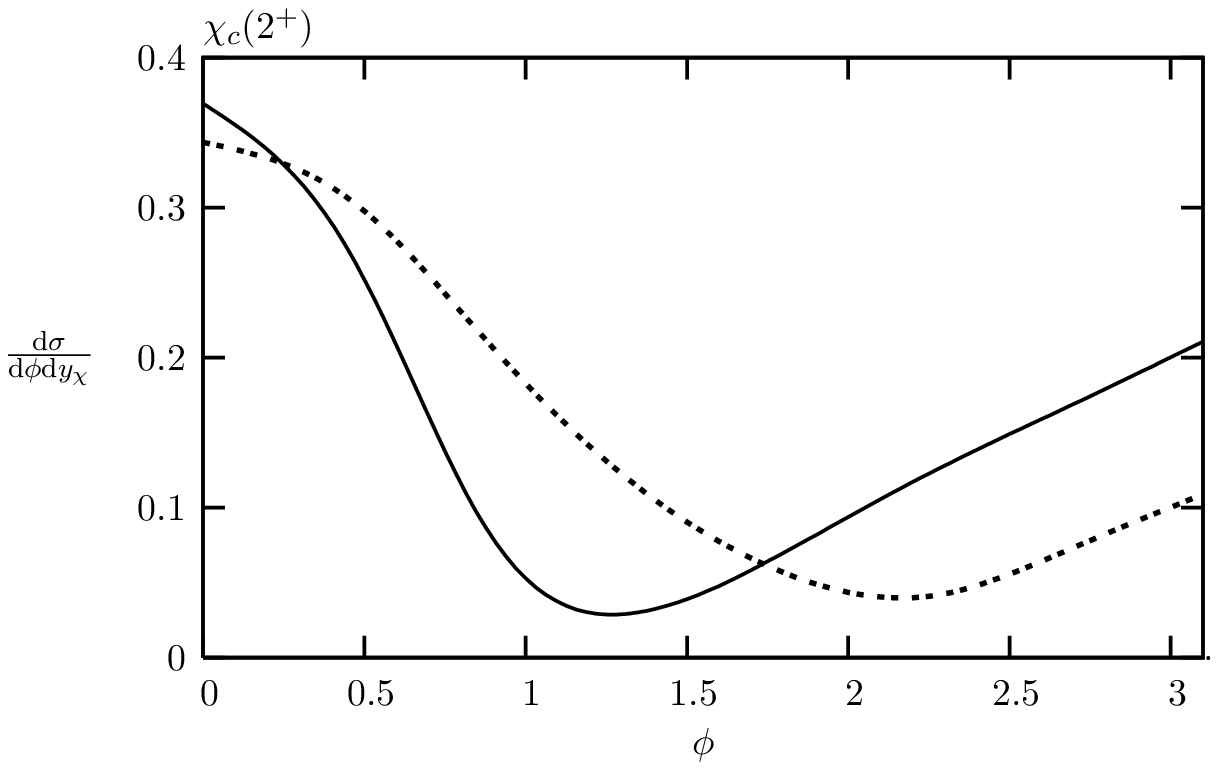}
\includegraphics[scale=0.5]{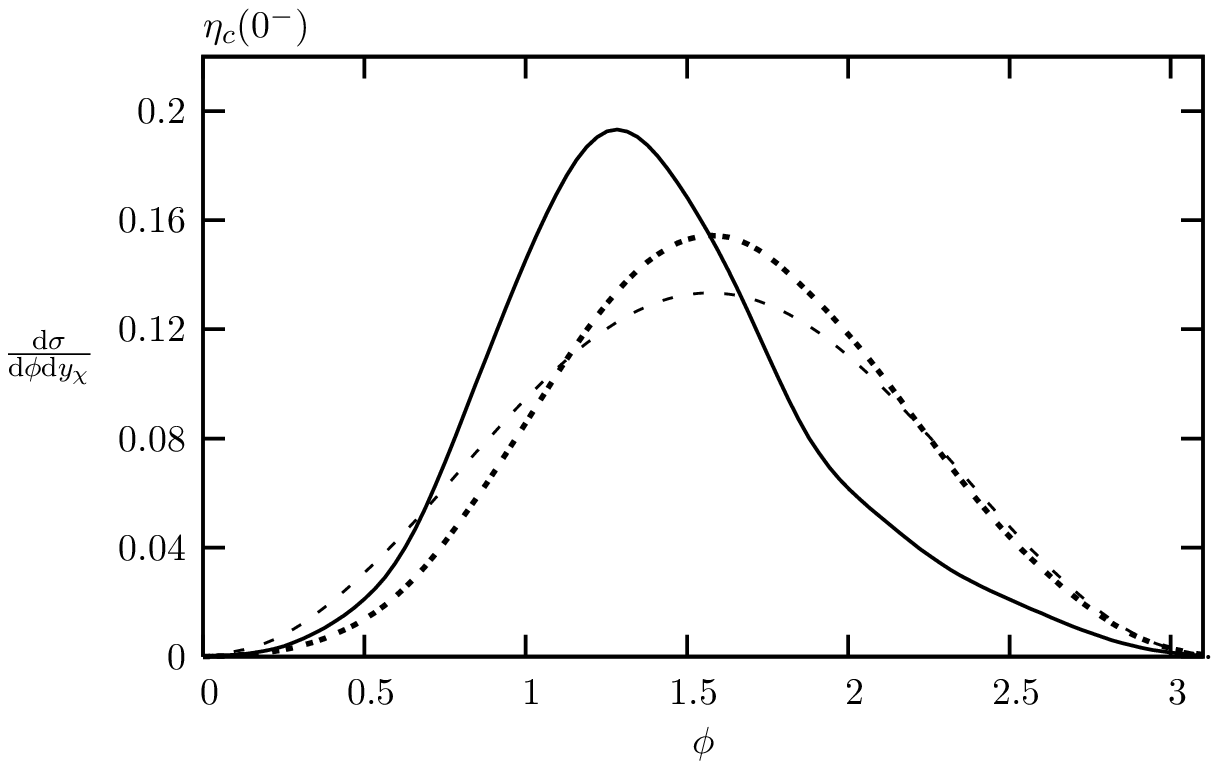}
\caption{Distribution (in arbitrary units) within the perturbative framework of the difference in azimuthal angle of the outgoing protons for the CEP of different $J^P$ $c\overline{c}$ states at $\sqrt{s}=14$ TeV and rapidity $y_X=0$. The solid (dotted) line shows the distribution including (excluding) the survival factor, calculated using the two channel eikonal model of Ref.~\cite{KMRsoft}, while the dashed line shows the distribution in the small $p_\perp$ limit, using the vertices of Eqs.~(\ref{R0p}--\ref{R0m}) and excluding the survival factor.}\label{surv1}
\end{center}
\end{figure}

As discussed in~\cite{Kaidalov03}, the CEP of different $J^P$ states results in characteristic distributions in $\phi$, the difference in azimuthal angle between the outgoing protons, and these are altered by absorptive corrections (which also depend on the particle spin and parity) resulting from multi-Pomeron exchanges. In Fig.~\ref{surv1} we show the ${\rm d}\sigma/{\rm d}\phi$ distribution for $\chi_{c(0,1,2)}$ and $\eta_c$ production at the LHC, while the following conclusions remains true for $\chi_b$/$\eta_b$ production and different c.m.s. energies, with the shape of the distributions only depending quite weakly on $\sqrt{s}$ and the central object mass $M_X$. The difference between the $J^P$ states, and the effect of including the soft survival factor, is clear. In Ref.~\cite{HarlandLang09} we demonstrated the difficulty of extracting spin information from the decay products of centrally produced $\chi_c$ states at the Tevatron, the basic problem being the low $p_\perp$ of the final state particles (which is characteristic of all CEP processes) which means that a sizeable fraction of the events that would allow spin discrimination will not pass the experimental $p_\perp$ cuts. Given that we expect the same issues to arise at the LHC, forward proton tagging would clearly provide an invaluable source of spin and parity information, and this is also true at RHIC, where we recall that forward proton detection is already possible.

We also see in Fig.~\ref{surv1} the effect of the $Q_\perp$ integral on the $\phi$ distributions. In the $\chi_0$ case we have a flat $\phi$ distribution as $p_\perp\to0$, but the inclusion of the $p_\perp$ dependent $gg \to \chi$ vertex factor and gluon propagators in (\ref{bt}) leads to corrections of the type $\sim{\bf p}_{1_\perp}\cdot{\bf p}_{2_\perp}/\langle Q_\perp^2\rangle$ which alter this. In the $\chi_1$ and $\eta$ cases, while the expected $\phi$ dependence is roughly the same as that predicted by (\ref{R1p}) and (\ref{R0m}), the $Q_\perp$ integral has again had some non-trivial effect on the original distributions. For $\chi_2$ production we can see that the $Q_\perp$ loop integral has induced a strong $\phi$ dependence, which we note cannot be predicted from general principles and is therefore specific to the perturbative model of CEP.

We have noted above that the inclusion of non-zero proton $p_\perp$ in (\ref{bt}) can be modelled by an `effective' slope $b_{\rm eff}$, and this was done in~\cite{HarlandLang09} to calculate the survival factors for central exclusive $\chi_{c(0,1,2)}$ production at the Tevatron. In particular we found that an increased survival factor can partly compensate the decrease in the `bare' $\chi_0$ cross section coming from the inclusion of non-zero $p_\perp$ effects, as well as changing the expected suppression of the $\chi_{1,2}$ CEP rates. However, we can see from Fig.~\ref{surv1} that the effect of including a non-zero proton $p_\perp$ is not just to change the slope $b$, but also to induce correlations between the outgoing proton momenta, which alter the $\phi$ distributions. Recalling that the survival probability is a function of $\phi$ as well as $p_\perp$ (through its dependence on the impact parameter ${\bf b}$), this may well have a non-trivial effect on the overall $p_\perp$ averaged suppression factor $\langle S^2_{\rm eik}\rangle$ which determines the final cross section. As an example of this we compare in Table~\ref{surv3} the value of the eikonal survival factor found by fitting the proton $p_\perp$ distribution with an effective slope $b_{\rm eff}$, with the result of the exact evaluation of (\ref{ampnew}), which includes the $\phi$ correlations present in Fig.~\ref{surv1}, for $\chi_c$/$\eta_c$ CEP at the Tevatron.\footnote{We note that these values are lower than those quoted in~\cite{HarlandLang09}, where the whole calculation was performed at $\sqrt{s}=60$ GeV to minimise PDF uncertainties, whereas we now fit the $p_\perp$ distributions and calculate the survival factors at the relevant collider energy, continuing to normalise relative to the $\chi_0$ cross section for $p_\perp=0$, calculated assuming a Regge extrapolation from the $\sqrt{s}=60$ GeV value. While the survival factor decreases, the `bare' cross section increases and the final predicted cross section is largely unchanged, but this procedure will give a more correct evaluation of the particle distributions.} While the values show an encouraging level of agreement (to within $\sim 10-20\%$), there is some difference between them, which is not surprising given the $\phi$ dependence of the cross section which fitting with $b_{\rm eff}$ omits. Moreover, in the case of $\chi_2$ production, for which we recall we cannot write an approximate closed form expression for the $gg \to \chi$ vertex as in Eqs.~(\ref{R0p}--\ref{R0m}), the only way to give a truly reliable estimate for the survival factor is by performing the integration of (\ref{ampnew}) exactly.\footnote{In Ref.~\cite{HarlandLang09} we assumed $|V_2|^2\sim p_{1_\perp}^2 p_{2_\perp}^2$ to calculate $S^2_{\rm eik}$ using the $b_{\rm eff}$ approximation. However, this assumption of a flat $\phi$ distribution is not really valid and in fact overestimates the expected soft suppression by a factor of $\sim 2$.} Table~\ref{surv4} lists the $p_\perp$ averaged survival factors, calculated using (\ref{ampnew}), for $\chi_c$ and $\eta_c$ production at RHIC, Tevatron and LHC energies; for $b\overline{b}$ production, the survival factors are in general slightly smaller, due to the larger value of $\langle {\bf Q}_\perp^2 \rangle$ and hence smaller $b_{\rm eff}$, but not significantly so.

Finally, although we have examined its limitations above, fitting the proton $p_\perp$ slope by $b_{\rm eff}$ (including now the effect of the survival factor via (\ref{rscr})) remains an effective way of modelling the expected $p_\perp$ distribution of the centrally produced particle as well as the proton $p_\perp$ distributions, integrated over $\phi$, in the MC. This is implemented for all of the processes discussed in this paper in the SuperCHIC Monte Carlo generator~\cite{SuperCHIC}, with the values for $\langle S_{\rm eik}^2\rangle$ calculated from (\ref{ampnew}), while a more complete inclusion of the angular correlations between the outgoing protons remains a possible future extension. In particular, in the light of the ongoing and future RHIC measurements with tagged forward protons~(see for example \cite{LeeDIS}), we plan to address this issue in a forthcoming publication~\cite{HKRSrhic}.

\begin{table}
\begin{center}
\begin{tabular}{|l|c|c|c|}
\hline
&$\chi_{c0}$&$\chi_{c1}$&$\eta_c$\\
\hline
$b_{\rm eff}$&0.068&0.16&0.18\\
exact &0.058&0.15&0.18\\
\hline
\end{tabular}
\caption{$p_\perp$ averaged survival factor $\langle S^2_{\rm eik}\rangle$ found using a $b_{\rm eff}$ fit and via the exact calculation of (\ref{ampnew}) for $\chi_{c(0,1)}$ and $\eta_c$ production at the Tevatron.}\label{surv3}
\end{center}
\end{table}

\begin{table}
\begin{center}
\begin{tabular}{|l|c|c|c|c|}
\hline
&$\chi_{c0}$&$\chi_{c1}$&$\chi_{c2}$&$\eta_c$ \\
\hline
Tevatron &0.058&0.15&0.11&0.18 \\
LHC (7~TeV) &0.037&0.11&0.084&0.13 \\
LHC (10~TeV) &0.033&0.10&0.078&0.11 \\
LHC (14~TeV) &0.029&0.091&0.072&0.10 \\
\cline{5-5}
RHIC &0.092&0.23&0.15& \multicolumn{1}{c}{}\\
\cline{1-4}
\end{tabular}
\caption{$p_\perp$ averaged survival factor $\langle S^2_{\rm eik}\rangle$ for $\chi_c$ and $\eta_c$ production at the Tevatron and LHC and $\chi_c$ production at RHIC ($\sqrt{s}=500$ GeV).}\label{surv4}
\end{center}
\end{table}
\subsection{Enhanced absorptive effects}\label{survenh}
Besides the effect of eikonal screening $S_{\rm eik}$, there is some suppression caused by the rescatterings of the intermediate partons (inside the unintegrated gluon distribution $f_g$). This effect is described by the so-called enhanced Reggeon diagrams and usually denoted as $S^2_{\rm enh}$, see Fig.~\ref{fig:pCp}. The value of $S^2_{\rm enh}$ depends mainly on the transverse momentum of the corresponding partons, that is on the argument $Q^2_i$ of 
$f_g(x,x',Q^2_i,\mu^2)$ in (\ref{bt}), and depends only weakly on the $p_\perp$ of the outgoing protons. While $S^2_{\rm enh}$ was previously calculated using the formalism of~\cite{Ryskin09}, we now use a newer version of this model~\cite{Ryskintba} which includes the continuous dependence on $Q^2_i$ 
and not only three `Pomeron components' with different `mean' $Q_i$. Thus we can now include the $S_{\rm enh}$ factor {\it inside} the integral (\ref{bt}), and so account for the dependence of $S^2_{\rm enh}$ on the integrand, which we recall will in general depend on the specific process being considered as well as the $x$ value being probed.

We show in Table~\ref{senhtab} the total suppression factor $\langle S^2_{\rm enh}\rangle$ for $\chi_{(c,b)0}$ CEP at RHIC, Tevatron and LHC energies (the values for the $\chi_{(1,2)}$ and $\eta$ states being almost the same as the respective $\chi_{0}$ value). As we can see, although the effect of including the enhanced absorptive corrections is already important at the Tevatron, the expected suppression is particularly large at LHC energies, with the size of the corrections increasing with the size of the available rapidity gaps $\sim\ln(s/M_X^2)$. Clearly, ignoring these corrections at the high values of $s/M_X^2$ relevant for low mass object CEP at the LHC will tend to overestimate the predicted cross section values by a sizeable amount (we note in particular that it is the inclusion of $S^2_{\rm enh}$ that largely explains the factor of $\sim 8$ decrease in the predicted $\chi_{c0}$ cross section relative to the result of~\cite{Khoze04}). Conversely, the observation of $\chi_c$, $\chi_b$ and $\gamma\gamma$ CEP at the LHC will allow a probing of $S^2_{\rm enh}$ at these record high values of $s/M_X^2$.
\begin{table}[ht]
\begin{center}
\begin{tabular}{|l|c|c|c|c|c|}
\hline
$\sqrt{s}$ (TeV)&0.5&1.96&7&10&14\\
\hline
$\chi_{c0}$&0.71&0.49&0.32&0.28&0.25\\
\hline
$\chi_{b0}$&--&0.65&0.43&0.37&0.32\\
\hline
\end{tabular}
\caption{Average $\langle S^2_{\rm enh}\rangle$, integrated over the gluon loop momentum ${\bf Q}_\perp$ and proton ${\bf p}_\perp$ for $\chi_{c0}$ and $\chi_{b0}$ production at different $\sqrt{s}$ values}\label{senhtab}
\end{center}
\end{table}
\subsection{Cross section results}
We calculate the expected heavy quarkonium CEP cross sections using the formalism of Section \ref{framew}. We use the $gg \to \chi/\eta$ vertices as given by Eqs.~(\ref{V0}--\ref{V0m}) but, consistently with (\ref{xcomp}), keep only the leading terms in $q_{i_\perp}^2/M^2$. As discussed in Sections \ref{surveff} and \ref{survenh}, we include the full $p_\perp$ dependence of the `eikonal' survival factor $S^2_{\rm eik}$ and the $gg \to X$ subprocess  amplitude in the calculation, as well as the $Q_\perp$ dependence of the `enhanced' survival factor $S^2_{\rm enh}$ in the $Q_\perp$ loop integral. 
The normalisation is set by the $\chi_0$ cross section calculated at $x=M_{\chi_{c0}}/(60~{\rm GeV})$ using GRV94HO partons~\cite{Gluck94}, as these extend to down to a very low scale ($Q_0^2\approx 0.4\,{\rm GeV}^2$), with a simple Regge scaling argument invoked to calculate the cross section at lower $x$ values, to minimise the significant PDF uncertainty at the low $x$ and $Q^2$ values we are considering here. In particular, the value of the $\chi_c$ cross section found by direct evaluation at Tevatron and especially LHC energies is strongly dependent upon the PDF set used, while at $\sqrt{s}=60$ GeV ($x\sim 0.05$) there is only a weak dependence, due to the much smaller PDF uncertainty at higher $x$--- see Section~\ref{gamres} for a further discussion of this in the context of $\gamma\gamma$ CEP. On the other hand, observables such as the ratios of the predicted perturbative $\chi_{c1,2}$, $\chi_b$ and $\eta_{(b,c)}$ cross sections relative to the predicted perturbative $\chi_{c0}$ cross section depend weakly on the PDF set used and so carry smaller overall uncertainties, although in particular the possibility of a sizeable non-perturbative contribution to the $\chi_{c(0,1,2)}$ cross section means it is difficult to quantify this statement. We would therefore argue that, as setting the normalisation at $\sqrt{s}=60$ GeV gives a predicted value for the Tevatron $\chi_c$ cross section that it is in good agreement with the data (see below), it should also give fairly reliable estimates for the heavy quarkonium CEP cross sections at RHIC, Tevatron and LHC energies, although the energy dependence of these cross section predictions, which depends on the gluon density $x$ dependence (recall that $\sigma \sim (xg)^4$) as well as soft survival effects, can only be estimated theoretically. 

A further uncertainty which is in particular worth recalling for the case of $c\overline{c}$ meson production is the size of the contribution to the cross section coming from the low $Q_\perp^2$ region of the integrand in (\ref{bt}), where perturbative QCD cannot really be trusted. Recalling that the infrared stability of the $Q_\perp$ integral depends on the presence of the hard mass scale $M_X$, it is not immediately clear that the $\chi$ mass is large enough to guarantee this. In particular, considering for simplicity the case of exactly forward proton scattering at $\sqrt{s}=60$ GeV, when the integral (\ref{bt}) is performed down to the GRV94H0 parton starting scale\footnote{We note that the contribution to the integral from the region $\Lambda_{\rm QCD}^2 \lesim Q_\perp^2<Q_0^2$ is negligible, irrespective of the precise procedure used to extrapolate the partons below $Q_0$} $Q^2=Q_0^2=0.4\,{\rm GeV}^2$ the expectation value $\langle Q_\perp^2 \rangle\sim 1\,{\rm GeV}^2$, with roughly half the contribution to the amplitude coming from the region below $Q_\perp=0.85$ GeV (see~\cite{Khoze04}). More realistically for $\sqrt{s}=1.96-14$ TeV we have $\langle Q_\perp^2 \rangle\sim 2-3\,{\rm GeV}^2$, due to the increase in $\partial\ln xg/\partial \ln Q^2$ (and therefore $\langle Q_\perp^2 \rangle$) at these lower $x$ values, with $\lesim 20\%$ of the amplitude coming from the region below $Q_\perp=0.85$ GeV, and so the situation is more encouraging. However the overall cross section normalisation at these low $x$ values depends sensitively on the largely uncertain PDFs-- see above.

Finally we recall that for $\chi_c$ CEP we also include a non-perturbative contribution to the cross section, calculated using a simple model where the Pomeron couples to each individual $c(\overline{c})$ quark (see Refs.~\cite{HarlandLang09,Khoze04} for a discussion of this as well as the details of the Regge scaling assumption used). In particular, in~\cite{Khoze04} the non-perturbative contribution at $\sqrt{s}=60$ GeV for forward scattering and prior to the inclusion of soft survival effects was found to be\footnote{We note that this cross section is normalised relative to $\Gamma(\chi_{c0} \to \gamma\gamma)$, extracted from the relevant branching ratio and the total width $\Gamma(\chi_{c0})$, taken from~\cite{PDG}. The non-perturbative contribution to the cross section prediction has therefore decreased by exactly the same factor of $\sim 1.4$, coming from the updated value for $\Gamma(\chi_{c0})$, as the perturbative contribution.}
\begin{equation}
\left.\frac{{\rm d}\sigma^{\rm nonpert}}{{\rm d}y_\chi}
\right|_{y_\chi=0}~\simeq~0.05~\mu{\rm b}\;,
\end{equation}
where we have integrated over the proton $p_\perp$ assuming the usual exponential form factor. This is then multiplied by the relevant survival factors, calculated for simplicity in the limit of exactly forward scattering (i.e. assuming that the only $p_\perp$ dependence of the non-perturbative amplitude comes from the proton form factor $\exp(bt)$), and can then be used to estimate the non-perturbative contribution to the $\chi_{c0}$ rate at higher $\sqrt{s}$ by assuming that the cross section exhibits a simple Regge scaling behaviour as in~\cite{Khoze04}. Explicitly, we find that the perturbative and non-perturbative contributions to the total $\chi_{c0}$ cross section are approximately equal ${\rm d}\sigma^{\rm non pert.} \approx {\rm d}\sigma^{\rm pert.}$ (see Table~\ref{chires2} for the total cross section values), roughly independent of the particular $\sqrt{s}$ value. This increase in the fractional non-perturbative contribution relative to the result of~\cite{Khoze04}, which found that ${\rm d}\sigma^{\rm non pert.} \approx {\rm d}\sigma^{\rm pert.}/2$, is due to the decrease in the perturbative cross section when a non-zero proton $p_\perp$ is included in the calculation. However, we note that the non-perturbative contribution, which was calculated in the forward limit of $p_\perp=0$, should also in principal be calculated using exact proton kinematics before a completely reliable comparison can be made. However, given the large uncertainty and model dependence in the calculation of the non-perturbative contribution, we will not consider a more detailed analysis of these issues here. We will also expect an equivalent non-perturbative contribution to the $\chi_{c(1,2)}$ and $\eta_c$ CEP rates which can in principal be calculated using the same model as in the $\chi_{c0}$ case. However, previous calculations~\cite{Peng95,Stein93} suggest that the non-perturbative contributions of the three $\chi_c$ states exhibit a similar hierarchy to the perturbative case (as given by (\ref{roughb})) so for simplicity we may assume, as in~\cite{HarlandLang09}, the same relative non-perturbative contribution to the $\chi_{c(1,2)}$ cross section as in the $\chi_{c0}$ case, and for consistency we make the same assumption for $\eta_c$ production. In the case of the $\chi_b$ and $\eta_b$ CEP the use of the perturbative framework is more justified on account of the large $M_X$ scale, and so we only include a perturbative contribution.

\begin{table}[ht]
\begin{center}
\begin{tabular}{|l|c|c|c|c|c|}
\hline
$\sqrt{s}$ (TeV)&0.5&1.96&7&10&14\\
\hline
$\frac{{\rm d}\sigma}{{\rm d}y_{\chi_c}}(pp\to pp(J/\psi\,+\,\gamma))$&0.57&0.73&0.89&0.94&1.0\\
\hline
$\frac{{\rm d}\sigma(1^+)}{{\rm d}\sigma(0^+)}$&0.59&0.61&0.69&0.69&0.71\\
\hline
$\frac{{\rm d}\sigma(2^+)}{{\rm d}\sigma(0^+)}$&0.21&0.22&0.23&0.23&0.23\\
\hline
\end{tabular}
\caption{Differential cross section (in nb) at rapidity $y_\chi=0$ for central exclusive $\chi_{cJ}$ production via the $\chi_{cJ} \to J/\psi\gamma$ decay chain, summed over the $J=0,1,2$ contributions, at RHIC, Tevatron and LHC energies, and calculated using GRV94HO partons, as explained in the text.}\label{chires1}
\end{center}
\end{table}
\begin{table}[ht]
\begin{center}
\begin{tabular}{|l|c|c|c|c|c|}
\hline
$\sqrt{s}$ (TeV)&0.5&1.96&7&10&14\\
\hline
$\frac{{\rm d}\sigma}{{\rm d}y_\chi}(\chi_{c0})$&27&35&42&43&45\\
\hline
$\frac{{\rm d}\sigma}{{\rm d}y_\chi}(\chi_{b0})$&--&0.017&0.021&0.022&0.022\\
\hline
\end{tabular}
\caption{Differential cross section (in nb) at rapidity $y_{\chi}=0$ for central exclusive $\chi_{(b,c)0}$ production at RHIC, Tevatron and LHC energies, and calculated using GRV94HO partons, as explained in the text.}\label{chires2}
\end{center}
\end{table}
\begin{table}[h]
\begin{center}
\begin{tabular}{|l|c|c|c|c|}
\hline
$\sqrt{s}$ (TeV)&1.96&7&10&14\\
\hline
$\frac{{\rm d}\sigma}{{\rm d}y_{\chi_b}}(pp\to pp(\Upsilon\,+\,\gamma))$&0.56&0.70&0.73&0.74\\
\hline
$\frac{{\rm d}\sigma(1^+)}{{\rm d}\sigma(0^+)}$&0.029&0.032&0.032&0.034\\
\hline
$\frac{{\rm d}\sigma(2^+)}{{\rm d}\sigma(0^+)}$&0.077&0.081&0.081&0.083\\
\hline
\end{tabular}
\caption{Differential cross section (in pb) at rapidity $y_\chi=0$ for central exclusive $\chi_{bJ}$ production via the $\chi_{bJ} \to \Upsilon\gamma$ decay chain, summed over the $J=0,1,2$ contributions, at Tevatron and LHC energies, and calculated using GRV94HO partons, as explained in the text.}\label{chires3}
\end{center}
\end{table}

\begin{table}[h]
\begin{center}
\begin{tabular}{|l|c|c|c|c|}
\hline
$\sqrt{s}$ (TeV)&1.96&7&10&14\\
\hline
$\frac{{\rm d}\sigma}{{\rm d}y_\eta}(\eta_c)$&200&200&190&190\\
\hline
$\frac{{\rm d}\sigma}{{\rm d}y_\eta}(\eta_b)$&0.15&0.14&0.14&0.12\\
\hline
\end{tabular}
\caption{Differential cross section (in pb) at rapidity $y_{\eta}=0$ for central exclusive $\eta_{b,c}$ production at Tevatron and LHC energies, and calculated using GRV94HO partons, as explained in the text.}\label{chires4}
\end{center}
\end{table}
In Table~\ref{chires1} we show predictions for the differential cross section for the central exclusive $pp\to pp(\chi_c) \to pp(J/\psi\gamma)$ process at RHIC, Tevatron and LHC energies. We note that, as in~\cite{HarlandLang09} (see also~\cite{teryaev,Pasechnik:2009qc}), a significant fraction of the $\chi_c$ events are predicted to correspond to the higher spin $\chi_{c(1,2)}$ states, although we note that the expected $\chi_{c(1,2)}$ contributions are now somewhat smaller than those quoted in~\cite{HarlandLang09}, due partly to the smaller values taken for the $gg \to \chi_{1,2}$~ K--factors (see Section~\ref{norm}) as well as, in the $\chi_{c2}$ case, the re-evaluated survival factor (as described in Section \ref{surveff}) and the larger value of $\langle Q_\perp^2 \rangle$ at higher $\sqrt{s}$ values which is now included (see footnote 6 of Section~\ref{surveff}). However, the precise values for their relative contributions should be taken as rough estimates only: as well as the overall uncertainty in the perturbative calculation at these low $M_\chi$ scales, higher order corrections to the $gg \to \chi$ vertices (which we recall are spin dependent) and the uncertainty in the spin-dependent soft survival factors, there is also the possibility of a non-perturbative contribution to the $\chi_{c(1,2)}$ rates to consider (see above).

Our prediction for the Tevatron is in good agreement with the CDF measurement of the $\chi_c$ central exclusive cross section~\cite{Aaltonen09chi}:
\begin{equation}
\frac{{\rm d}\sigma}{{\rm d}y_{\chi_c}}(pp\to pp(J/\psi\gamma))\bigg|_{y_\chi=0}=(0.97\pm 0.20)\;{\rm nb}\;.
\end{equation}
Given the large uncertainty in the cross section prediction (which we roughly estimate to be of order $\sim {}^{\times}_{\div} 5$ -- see Section 3 of~\cite{HarlandLang09} for a more detailed discussion of these uncertainties) for the low mass $\chi_c$ states, the closeness of this agreement is not necessarily to be expected, although it lends broad support to the calculation. On the other hand, this uncertainty is significantly reduced when we consider the ratios of the $\chi_{c0}$ to $\chi_{c(1,2)}$ perturbative contributions: our results therefore suggest that a non-negligible fraction of the observed $\chi_c$ events at the Tevatron are in fact $\chi_{c1}$ and $\chi_{c2}$ events, although a precise prediction of their relative contributions is difficult to make (we recall in particular the uncertainty in the spin-dependent NLO corrections to the $gg \to \chi_J$ vertices, as discussed in Section~\ref{norm}). For clarity, we also show in Table~\ref{chires2} predictions for the $\chi_{c0}$ (and $\chi_{b0}$) CEP cross sections, which would be relevant for the observation of $\chi_c$ CEP via two body decay channels (e.g. $\chi_c \to \pi\pi,K\overline{K}$) -- see the Conclusion section for a more detailed discussion of this.

In Table~\ref{chires3} we show predictions for the differential cross section for the central exclusive $pp\to pp(\chi_b) \to pp(\Upsilon\gamma)$ process at Tevatron and LHC energies. While the overall rate is greatly reduced by the strong $M_\chi$ dependence of the Sudakov factor (\ref{ts}), $\chi_b$ CEP remains a potential observable at the LHC. We can see, as discussed in Section \ref{norm}, that $\chi_{b1}$ states will give a negligible contribution to the overall rate, while the relative $\chi_{b2}/\chi_{b0}$ contribution is reduced in comparison to the $\chi_c$ case. In fact, this suppression is largely due to our choice of branching ratio ${\rm Br}(\chi_{b0}\to\Upsilon\gamma)\approx 3 \%$: clearly, if we took a smaller value this would increase the expected relative $\chi_{b2}/\chi_{b0}$ contribution. Nevertheless, we can predict with some certainty that the $\chi_{b0}$ contribution to any future observed $\chi_b$ events (via the $\chi_b \to \Upsilon\gamma$ decay chain) will be strongly dominant.

In both cases, we can see that the predicted $\chi_{(c,b)}$ CEP cross sections depend only weakly on the c.m.s. energy. While the higher gluon density at lower $x$ will lead to an increase in the cross section, this growth is tamed by the eikonal and enhanced soft survival factors, which decrease with $\sqrt{s}$ due to the increase in proton opacity $\Omega(s,b)$ and increase in the size of the rapidity gaps $\sim \ln(s/M_X^2)$ available for `enhanced' absorption, respectively. Clearly the exact energy dependence of both the gluon parton density and the soft survival factors carry their own uncertainties and as a consequence the predicted energy dependence of the CEP rates should be considered as an estimate only. However this weak dependence still represents a qualitative prediction, the validity of which could be probed by observations of these processes at RHIC, Tevatron and/or different LHC running energies. In particular, we note that the predicted RHIC $\chi_c$ and $\chi_{c0}$ cross sections are not greatly reduced relative to the Tevatron values and $\chi_c$ CEP would therefore certainly represent a potential observable at RHIC. Moreover, although we do not consider it explicitly here, we note that the expected $\chi_{b0}$ cross section at RHIC is also predicted to be comparable to the predicted Tevatron rate.

Finally, we show in Table~\ref{chires4} predictions for the differential cross section for central exclusive $\eta_c$ and $\eta_b$ production at Tevatron and LHC energies. In both cases, the expected rates are roughly two orders of magnitude smaller than the associated $\chi_{c,b}$ cross sections, consistent
 with the expected $\sim \langle \mathbf{p}_{\perp}^2\rangle^2/(2\langle \mathbf{Q}_{\perp}^2\rangle^2)$ suppression described in Section \ref{vertex}. We can also see that the cross sections are approximately flat/slowly decreasing with energy: in the case of $\eta$ CEP the amplitude (\ref{bt}) is more strongly dependent on low values of $Q_\perp$ than in the $\chi_0$ case (as $Q^2_\perp$ in the numerator is replaced by $p_\perp^2$), and at lower $Q$ scales the PDFs increase more slowly with decreasing $x$.

\section{Central exclusive $\gamma\gamma$ production}\label{gamCEP}
Central exclusive diphoton production was first considered in~\cite{Khoze04gg}, with the cross section written in the factorised form~\cite{KMRprosp}
\begin{equation}\label{fact}
\sigma=\mathcal{L}_g(M_{X}^2,y_X)\hat{\sigma}(M_X) ,
\end{equation}
where $\hat{\sigma}$ is the cross section for the hard $gg \to \gamma\gamma$ subprocess, which proceeds via an intermediate quark loop, and $\mathcal{L}_g$ is the effective $gg$ luminosity for the production of a central system $X$ ($=\gamma\gamma$ here) of mass $M_X$ and rapidity $y_X$. To single log accuracy this is given by
\begin{align}
\frac{\partial\,\mathcal{L}_g}{\partial y_X\partial\,{\rm ln}M_X^2}=S^2_{\rm eik}\bigg(\frac{\pi}{(N_C-1)^2b}\int\frac{{\rm d}Q_\perp^2}{Q_\perp^4}
f_g(x_1,x_1',Q_\perp^2,\mu^2)f_g(x_2,x_2',Q_\perp^2,\mu^2)\bigg)^2\;,
\end{align}
where $S^2_{\rm eik}$ is the eikonal survival factor (enhanced rescattering was not considered), and the factor of $1/b^2$ comes from the integral over the exponential proton form factor, with the outgoing proton $p_\perp$ neglected in the $f_g$ and $\hat{\sigma}$ calculation. Thus all of the physics coming from the Sudakov factor and skewed gluon PDFs is factored off in a universal luminosity function that does not depend on the specific subprocess cross section $\hat{\sigma}$. Such a formulation can readily be shown to follow from (\ref{bt}) in the limit that ${\mathbf p}_{1_\perp}={\mathbf p}_{2_\perp}=0$, but in general the ${f_g}$ functions, the survival factor $S^2_{\rm eik}(b_t)$, and hard subprocess amplitude $\hat{\sigma}$ will depend on the outgoing proton $p_\perp$ and so the simple multiplicative factorisation of (\ref{fact}) will not hold. The inclusion of these effects will not only in general lead to a more reliable cross section estimate in the $J_z=0$ case, but also allows for the inclusion of contributions that violate the $J_z=0$ selection rule~\cite{Khoze00a}, which is only exact in the limit that the proton $p_\perp=0$. In the case of central exclusive $\chi_c$ production, for example, these have been observed to have a significant effect~\cite{HarlandLang09,Pasechnik:2009qc}.
\subsection{$|J_z|=2$ contribution}
\begin{figure}
\begin{center}
\includegraphics[scale=0.7]{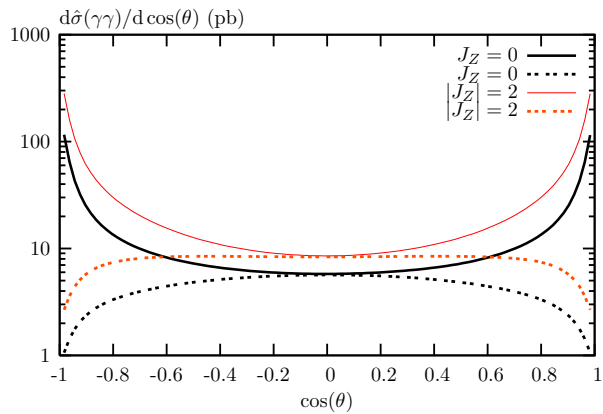}
\caption{Centre-of-mass scattering angle dependence of the hard subprocess $gg\to\gamma\gamma$ cross section, averaged over incoming gluon polarizations at the amplitude level, for a $J_z=0$ and $|J_z|=2$ incoming $gg$ system. The continuous curve represents production for a fixed $M_{\gamma\gamma}=10$~GeV, and the dashed for fixed $E_{\perp_\gamma}=5$~GeV.}\label{gamsig}
\end{center}
\end{figure}

Taking (\ref{bt}) as our starting equation, to separate out the different $J_z$ contributions we must decompose the sum $q_{1_\perp}^\mu q_{2_\perp}^\nu V_{\mu\nu}$ of (\ref{Vnorm}) in terms of the incoming gluon polarization vectors, given by
\begin{align}\nonumber
\epsilon^{_{+(-)}}_{_{1(2)}}&=-\frac{1}{\sqrt{2}}(\hat{x}+i\hat{y})\,\\ \label{pol}
\epsilon^{_{-(+)}}_{_{1(2)}}&=\frac{1}{\sqrt{2}}(\hat{x}-i\hat{y})\,
\end{align}
where the $x-y$ plane is perpendicular to the direction of motion of the gluons in the $gg$ rest frame, which in the on-shell approximation (valid up to corrections of order $\sim q_\perp^2/M_{\gamma\gamma}^2$) coincides with the transverse plane in the lab frame. We can then invert (\ref{pol}) to change the incoming momenta vectors $q_\perp$ to the helicity basis, giving
\begin{align}
q_{1_\perp}^i q_{2_\perp}^j \mathcal{M}_{ij} =\begin{cases} &-\frac{1}{2} ({\bf q}_{1_\perp}\cdot {\bf q}_{2_\perp})(\mathcal{M}_{++}+\mathcal{M}_{--})\;\;(J^P_z=0^+)\\ 
&-\frac{i}{2} |({\bf q}_{1_\perp}\times {\bf q}_{2_\perp})|(\mathcal{M}_{++}-\mathcal{M}_{--})\;\;(J^P_z=0^-)\\ 
&+\frac{1}{2}((q_{1_\perp}^x q_{2_\perp}^x-q_{1_\perp}^y q_{2_\perp}^y)+i(q_{1_\perp}^x q_{2_\perp}^y+q_{1_\perp}^y q_{2_\perp}^x))\mathcal{M}_{-+}\;\;(J^P_z=+2^+)\\ 
&+\frac{1}{2}((q_{1_\perp}^x q_{2_\perp}^x-q_{1_\perp}^y q_{2_\perp}^y)-i(q_{1_\perp}^x q_{2_\perp}^y+q_{1_\perp}^y q_{2_\perp}^x))\mathcal{M}_{+-}\;\;(J^P_z=-2^+)
\end{cases}\label{Agen}
\end{align}
where $\mathcal{M}_{\lambda_1\lambda_2}$ are the $g(\lambda_1)g(\lambda_2)\to \gamma(\lambda_3)\gamma(\lambda_4)$ helicity amplitudes (with the photon helicity labels implicit), which can be found in the literature, see for example Ref.~\cite{Bern:2001dg}. We first note that in the $p_\perp\to 0$ limit the only non-vanishing term after the $Q_\perp$ integration is the first one, with
\begin{equation}
q_{1_\perp}^i q_{2_\perp}^j \mathcal{M}_{ij} \to \frac{1}{2}Q_\perp^2(\mathcal{M}_{++}+ \mathcal{M}_{--})\sim\sum_{\lambda_1,\lambda_2}\delta^{\lambda_1\lambda_2}\mathcal{M}_{\lambda_1\lambda_2}\;,
\end{equation}
i.e. the expected $J_z^P=0^+$ selection rule. Concentrating on the $J_z=\pm 2$ piece, we note that this can be written in the manifestly covariant form
\begin{equation}
\epsilon_{\mu\nu}^{(+2)} q_{1_\perp}^\mu q_{2_\perp}^\nu\epsilon^{_-}_1\epsilon^{_+}_2 \mathcal{M}_{-+}+\epsilon_{\mu\nu}^{(-2)} q_{1_\perp}^\mu q_{2_\perp}^\nu\epsilon^{_+}_1\epsilon^{_-}_2 \mathcal{M}_{+-}\;,
\end{equation}
where the $\epsilon_{\mu\nu}^{(\pm 2)}$ are the usual $|J_z|=2$ polarization tensors (evaluated in the rest frame of the $gg$ system). Noting that Bose symmetry gives $\mathcal{M}_{\lambda_1,\lambda_2}(s,t,u)=\mathcal{M}_{{\lambda_2,\lambda_1}}(s,u,t)$, we can give a rough estimate for the expected contribution by considering for example the production of a $(++)$ $\gamma\gamma$ state, for which we have the explicit LO result $\mathcal{M}_{\lambda_1\lambda_2\lambda_3\lambda_4}=\mathcal{M}_{+++-}=1$~\cite{Bern:2001dg}. In this case the $|J_z|=2$ contribution simplifies to
\begin{equation}\label{qjz2}
T(|J_z|=2) \sim (q_{1_\perp}^x q_{2_\perp}^x-q_{1_\perp}^y q_{2_\perp}^y) \;.
\end{equation}
After performing the $Q_\perp$ integral and squaring, this will be of order
\begin{equation}\label{simjz2}
|T|^2 \sim p_{1_\perp}^2p_{2_\perp}^2 \to \frac{\langle p_\perp^2 \rangle^2}{\langle Q_\perp^2\rangle^2}\;,
\end{equation}
i.e. with the same level of suppression as in the $\chi_2$ CEP case, as we would expect. While the $|J_z|=2$ subprocess cross section $\hat{\sigma}$ is a factor of $\sim 2$ larger than the $J_z=0$ cross section, this level of suppression is quite significant, and an explicit calculation of the full contribution, summed over the final state photon polarizations, gives
\begin{equation}
\frac{|T(|J_z|=2)|^2}{|T(J_z=0)|^2} \sim 1\%
\end{equation}
with some variation depending on the particular PDF set used and the $x$ value considered. While this will receive some compensation from a larger survival factor, as in the $\chi_{c2}$ case, this suppression is further increased when the $|\eta_\gamma|$ cuts are included, as the average $\langle |\cos\theta| \rangle$ is somewhat larger for the $|J_z|=2$ bare cross section.  Moreover, in Fig.~\ref{gamsig} we plot the $J_z=0$ and $|J_z|=2$ subprocess differential  cross sections, and we can see that the angular distributions for the two spin cases do not differ significantly, as they are dominated in both cases by double logarithmic singularities in the amplitudes as $u,t \to 0$. The $|J_z|=2$ contribution to $\gamma\gamma$ CEP is therefore negligible, while similar considerations show the $0^-$ contribution in (\ref{Agen}) to be further suppressed. We will therefore only consider the $J_z^P=0^+$ contribution to the $\gamma\gamma$ CEP cross section.
\subsection{Exclusive $\pi^0\pi^0$ background}

An important possible background to $\gamma\gamma$ CEP is the exclusive production of a pair of $\pi^0$ mesons, with one photon from each $\pi^0$ decay undetected or the two photons merging~\cite{CDFgg}. At first sight it would appear that the cross section for this purely QCD process may be much larger than the $\gamma\gamma$ cross section and so would constitute an appreciable background, but fortunately this is not the case. First, using purely dimensional arguments we can see that the amplitude to form an exclusive pion with large transverse energy will be proportional to the ratio $f_\pi/E_\perp$, that is the cross section of the $gg^{PP}\to \pi^0\pi^0$ hard subprocess contains the numerically small factor $(f_\pi/E_\perp)^4$ which in the region of interest is comparable to (or even smaller than) the QED suppression, $(4\pi\alpha_{QED})^2$, of the $gg^{PP}\to\gamma\gamma$ cross section. 

Secondly, as we shall discuss in detail in a future publication~\cite{Harlandlangfut}, the LO amplitude for $\pi^0\pi^0$ exclusive production in a $J_z=0$ state vanishes in the same way as the $\gamma\gamma\to \pi^0\pi^0$ amplitude \cite{Brodsky81,Chernyak06}. Thus the incoming active gluons $gg^{PP}$ will principally be in an admixture of the $|J_z|=2$ state, which we recall is strongly suppressed (by a factor of $\sim (\langle p^2_\perp\rangle/\langle Q^2_\perp\rangle)^2$ (\ref{simjz2})) due to the $J_z=0$ selection rule which operates for forward outgoing protons. Therefore we can safely conclude that exclusive $\pi^0$ pair production will not constitute a large background to the central exclusive $pp\to p+\gamma\gamma+p$ process, even before any consideration of the efficiency with which $\pi^0 \to \gamma\gamma$ mimics single $\gamma$ production.

\subsection{Cross section calculation and results}\label{gamres}
We calculate the $J_z^P=0^+$ contribution to the $\gamma\gamma$ CEP cross section using (\ref{bt}), with the helicity amplitudes for the hard $gg\to\gamma\gamma$ subprocess given in Ref.~\cite{Bern:2001dg}. An explicit calculation shows that the inclusion of non-zero proton $p_\perp$ in the hard subprocess calculation leads to a factor $\sim2$ decrease in the `bare' cross section (prior to the inclusion of soft survival effects) with some small variation depending on the specific cuts imposed, the central mass $M_{\gamma\gamma}$ (in particular at very low mass), and the PDF set used. However such a decrease is typically associated with an increase in the steepness of the slope $b_{\rm eff}$ of the proton form factor, with the exact amount in principle being a function of the central system mass $M_{\gamma\gamma}$. The result of this increased $b_{\rm eff}$ value is that the reaction has now become more peripheral, leading to an increase in the eikonal survival factor $S^2_{\rm eik}$ which partly compensates the initial decrease in the cross section, with the exact amount calculated as in Eq.~(\ref{ampnew}) of Section~\ref{surveff}. In fact, as $M_X$ is increased we expect the saddle point $\langle Q_\perp^2 \rangle$ of the integrand of (\ref{bt}), which we recall largely determines the value of $b_{\rm eff}$, to increase due to the higher scale of the Sudakov factor, but as we are also probing higher $x$ values we will find a decrease in $\partial{\rm ln}\,xg/\partial{\rm ln}\,Q^2$, which will tend to reduce $\langle Q_\perp^2 \rangle$. These effects largely cancel out, and as a result the value of $b_{\rm eff}$ only depends weakly on $M_{\gamma\gamma}$. The net effect is a factor of $\approx 30\%$ decrease in the cross section which is roughly constant with $M_{\gamma\gamma}$ and a moderately steeper proton $p_\perp$ distribution with slope $b_{\rm eff} \sim 5\,{\rm GeV}^{-2}$, which is further increased to about $b_{\rm exp} \sim 6\,{\rm GeV}^{-2}$ upon the inclusion of soft survival effects (see Section~\ref{surveff}).

\begin{figure}[t]
\begin{center}
\includegraphics[scale=0.55]{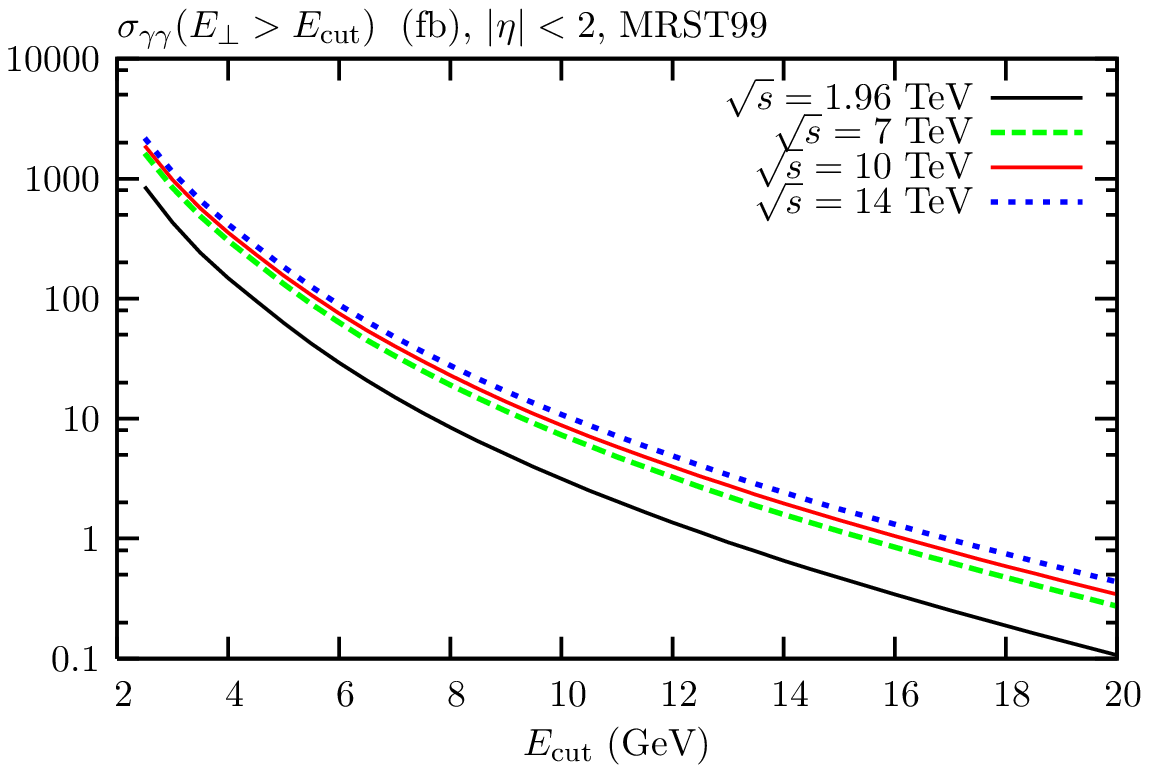}
\includegraphics[scale=0.55]{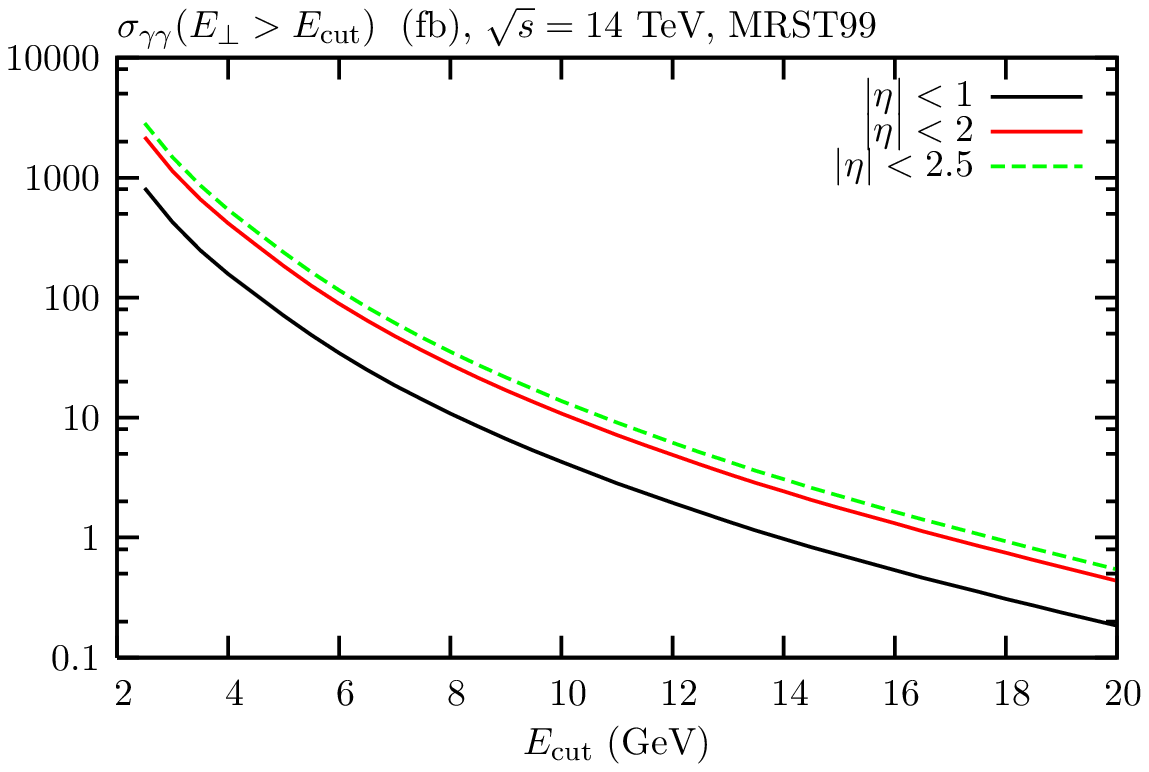}
\caption{Central exclusive $\gamma\gamma$ cross section at different $\sqrt{s}$ values, and for different values of $\eta_{\rm max}$, with the emitted photons required to have $E_\perp>E_{\rm cut}$.}\label{gam1}
\end{center}
\end{figure}
We provide two sets of predictions using the MRST99~\cite{Martin99} and MSTW08LO~\cite{Martin09} partons sets at c.m.s. energies of $\sqrt{s}=$1.96, 7, 10, 14~TeV for a range of photon pseudorapidities $\eta_{\gamma}$, and transverse energy $E_{\perp}$ values, calculated using the SuperCHIC Monte Carlo generator. In Fig.~\ref{gam1} we show the predicted cross section at these c.m.s. energies as a function of the cut, $E_{\rm cut}$, on the photon transverse energy, $E_\perp$. As we have $M_{\gamma\gamma}/2 \sim E_\perp \sim E_{\rm cut}$, this cut effectively controls the invariant mass of the $\gamma\gamma$ system being produced. The steep fall-off with $M_{\gamma\gamma}$ coming from the Sudakov factor, the shape of which is largely independent of the overall uncertainties of the calculation, is clear, and given enough data this distribution can certainly be tested. We also show in Fig.~\ref{gam1} the effect of changing the cut on the photon pseudorapidity $\eta_{\gamma}$ at $\sqrt{s}=14$~TeV. As we would expect there is quite a large increase in the rate as higher photon $\eta_{\gamma}$ values are accepted. 

\begin{figure}[t]
\begin{center}
\includegraphics[scale=0.55]{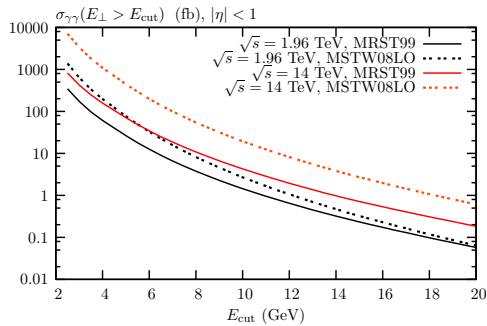}
\caption{Comparison of central exclusive $\gamma\gamma$ cross section at $\sqrt{s}=1.96,\,14$ TeV using MRST99 and MSTW08LO partons}\label{gam2}
\end{center}
\end{figure}
For the lower $M_X$ values it is important to emphasise the large degree of uncertainty in the single PDFs at the low $x$ and $Q^2$ values we are considering (see for example Fig.~5 of~\cite{HarlandLang09}). Recalling the strong PDF dependence ($\sigma_{\rm CEP} \sim (xg)^4$) of the CEP cross section, the effect of this will be quite severe. This is already relevant at Tevatron energies, where the typical $x$ value considered here is of order $\sim 0.001-0.005$, but is a stronger source of uncertainty when it comes to making predictions for the LHC, where the probed $x$ values are even lower. In Fig.~\ref{gam2} we show this explicitly by comparing the predicted CEP rates at LHC and Tevatron energies using the MRST99 and MSTW08LO sets. Although there is some convergence at Tevatron energies between the two sets as $M_{\gamma\gamma}$ is increased, the discrepancy at $\sqrt{s}=14$ TeV is clearly very large across all realistic $M_{\gamma\gamma}$ values, while the situation is only mildly improved at the lower $\sqrt{s}=7,10$ TeV values. We use these two particular sets because we would argue that they span a realistic range of small $x$ parton distributions. MSTW08LO is the outcome of a leading-order pQCD global fit to an up-to-date set of DIS and other hard scattering data. At small $x$, it gives a reasonable, though not perfect, description of HERA $F_2(x,Q^2)$ data. The corresponding NLO version (MSTW08NLO) has a very different gluon distribution at small $x$ and $Q^2$, with $g(x,Q_0^2) < 0$ for $ x \lesim 10^{-2}$. Because of this behaviour, the MSTW08NLO gluon gives unstable results when used in (\ref{fgskew}) to calculate $f_g$. We prefer to use the older MRST99 NLO set, which has a more benign small-$x$ form, while still retaining the essential features of a NLO fit, in particular a gluon that is smaller at small $x$ than at LO. We can therefore regard the MSTW08LO and MRST99NLO sets as providing approximate upper and lower bounds, respectively, on the range of predictions coming from different PDF sets. With this in mind, we show in Tables~\ref{tevtab} and~\ref{lhctab} the cross section predictions using these two PDF sets at Tevatron and LHC energies, respectively, for a range of cuts on the photon $E_\perp$ and central mass $M_{\gamma\gamma}$, with the experimentally most relevant pseudorapidity cuts $|\eta_\gamma|<1.8$ (Tevatron) and $|\eta_\gamma|<2$ (LHC) imposed~\cite{Albrowpriv1}.
\begin{table}
\begin{center}
\begin{tabular}{|l|c|c|c|c|c|}
\hline
$E_{\rm cut}$& MRST99 & MSTW08LO &$M_{\rm min}$& MRST99 & MSTW08LO\\
\hline
2 & 1790 & 8000 &4&3870&18900 \\
5 & 57.1 & 163 &10&114&345 \\
10 & 2.86 & 5.43&20&5.65&11.36  \\
\hline
\end{tabular}\caption{Central exclusive $\gamma\gamma$ production cross sections (in fb) at the Tevatron for different values of cuts on the $E_\perp$ ($>E_{\rm cut}$) of the final-state photons and the invariant mass $M_X$ ($>M_{\rm min}$) of the diphoton system, in GeV. The photons are restricted to lie in the centre of mass rapidity interval $|\eta_\gamma|<$1.8.}\label{tevtab}
\end{center}
\end{table}

\begin{table}
\begin{center}
\begin{tabular}{|l|c|c|c|c|c|c|}
\hline
&$E_{\rm cut}$& MRST99 & MSTW08LO &$M_{\rm min}$& MRST99 & MSTW08LO\\
\hline
$\sqrt{s}=7$ TeV&5&126&596&10&264&1321 \\
&10&6.58&22.6&20&13.8&49.4 \\
&15&0.988&2.82&30&2.14&6.35  \\
&20&0.226&0.567&40&0.524&1.35  \\
\hline
$\sqrt{s}=10$ TeV &5&147&800&10&306&1770 \\
&10&7.77&31.0&20&16.2&67.9 \\
&15&1.20&3.94&30&2.59&8.83  \\
&20&0.274&0.806&40&0.632&1.91  \\
\hline
$\sqrt{s}=14$ TeV &5&172&1066&10&358&2340 \\
&10&9.46&42.7&20&19.7&92.8 \\
&15&1.46&5.54&30&3.14&12.4  \\
&20&0.344&1.15&40&0.792&2.71  \\
\hline
\end{tabular}
\caption{Central exclusive $\gamma\gamma$ production cross sections (in fb) at different LHC c.m.s energies for different values of cuts on the $E_\perp$ ($>E_{\rm cut}$) of the final-state photons and the invariant mass $M_X$ ($>M_{\rm min}$) of the diphoton system, in GeV. The photons are restricted to lie in the centre of mass rapidity interval $|\eta_\gamma|<$2.}\label{lhctab}
\end{center}
\end{table}

Although it is evidently difficult to make precise predictions for the LHC, there are a number of ways in which the uncertainty can be decreased. First, the Tevatron $\gamma\gamma$ cross section, which has been measured, or indeed early LHC observations at lower $\sqrt{s}$ values may be used as rough `benchmark' normalisation points with which we can select the most appropriate PDF set. Clearly there are other quite large uncertainties in the predicted Tevatron $\gamma\gamma$ cross section, in particular due to the soft survival factors (the energy dependence of which is also not known precisely), and moreover the difference between the PDF set predictions increases as we go to higher energies, so such a procedure can only be quite approximate but may nevertheless be useful. A second related point to note is that the PDF uncertainties present in the cross section normalisations are reduced to roughly a factor of $\lesim 2$ when instead cross section ratios at different $\sqrt{s}$ values are considered. Lastly, we note that the MRST99 partons, which exhibit an approximate Regge-like flat low $x$ dependence at low $Q^2$, predict a $\chi_c$ CEP cross section that is in better agreement with the observed Tevatron value~\cite{Aaltonen09} (in particular, direct calculation at $\sqrt{s}=1.96$ TeV gives a cross section that is roughly a factor $\sim 1.5$ higher than the result calculated in~\cite{Khoze04}), and so perhaps provide a more reliable estimate of the expect low mass CEP cross sections we are considering here, although given the theoretical uncertainties in the $\chi_c$ calculation this can only be taken as a rough guide.

\section{Conclusions}\label{conc}

In this paper we have presented predictions for a number of important standard candle CEP processes at Tevatron, RHIC and LHC energies. We have focused on heavy $(c,b)$ $\chi$ and $\eta$ quarkonia and $\gamma\gamma$ production, significantly improving and extending our previous studies in a number of ways. We have considered the uncertainties in the calculations due to poorly determined quarkonium bound-state properties, from unknown higher-order perturbative corrections, from possible non-perturbative contributions, and from the behaviour of the skewed unintegrated gluon density at small $x$. Improvements in the calculation include a more careful treatment of enhanced absorptive effects, which can have a significant effect on the overall production cross sections. We have presented heavy quarkonia cross section predictions for a variety of collider energies, and also considered the distributions in the relative azimuthal angle of the outgoing protons, which can in principle provide complementary information on the properties of the survival factors.

Central exclusive $\gamma\gamma$ production is another important standard candle process, and indeed has already been observed at the Tevatron. A careful treatment of the various angular momentum states shows that the $J_z^P = 0^+$ contribution is expected to dominate. We have presented predictions for cross sections and photon transverse momentum distributions at the Tevatron and the LHC, again highlighting the uncertainties in the calculation coming from the small-$x$ PDFs. A precise measurement of $\gamma\gamma$ CEP at the Tevatron will be important in benchmarking these calculations, and should reduce the uncertainties in the LHC predictions. We have also shown that the background from CEP $\pi^0\pi^0$  production is expected to be small. 

What are the prospects for CEP measurements at the LHC? Note that at the moment, without forward proton taggers, the existing LHC detectors (ALICE, ATLAS, CMS and LHCb) are not well suited for studying CEP processes as they lack  the coverage necessary to measure large rapidity gaps. Selecting only events with a low multiplicity in the central detector, corresponding to $\chi\to \pi\pi$, $\chi\to KK$ and $\chi\to J/\psi\gamma$ decays, may not be enough to study CEP since such events
include processes with incoming proton dissociation: in the case of high mass dissociation there is no $J_z=0$ selecting rule,
characteristic of CEP processes. However, as discussed in \cite{FSC,jerry}, the addition of forward shower counters
(FSC) along the beam line would provide a record rapidity coverage and would allow the exclusion of events with high mass and a large fraction of events with low
mass diffractive dissociation. Thus events with a `veto' FSC trigger\footnote{It would be better to use the `veto' FSC trigger at relatively low luminosities.
When the mean number of inelastic $pp$ interactions per bunch crossing $n\gg 1$ becomes large, the efficiency of the FSC `veto' trigger decreases as  $e^{-n}$ since the FSC will be filled by secondaries from the `pile-up' events. Therefore an `effective' luminosity for selected CEP
events, $L_{\rm eff}=L_0e^{-n}$, will be much smaller than the true LHC luminosity $L_0$. The rich physics venues which FSCs will provide in studies of CEP processes may therefore be considered as an additional strong argument in favour of their installation at the LHC as soon as possible.} (that is without the signal in FSC) may be considered as CEP process\footnote{The cross section will be a bit larger due to the admixture of a low mass, $N\to N^*$, dissocition but this will not affected the main qualitative and quantitative features of the process.} and can be used to study different diffractive reactions. It is important to emphasise that, as discussed in \cite{HarlandLang09,Khoze04}, in the absence of forward proton detectors it would also be instructive to observe central exclusive $\chi_c$ production
via the two-body decay modes, in particular the {\it spin-analyzing} $\pi\pi$ and $K\bar{K}$ channels. We recall that both  the $\pi\pi$ and $K\bar{K}$ decay modes of the $\chi_{c0}$ meson have a branching fraction of about 1$\%$, while these decay channels are forbidden for $\chi_{c1}$  and suppressed by about a factor of 5 for the $\chi_{c2}$ relative to the $\chi_{c0}$. Another promising mode  is $\chi_c\to p\bar{p}$, since the branching fraction for $\chi_{c0}$ ($\simeq 0.024\%$) is a factor of 3 higher than that for $\chi_{c1,2}$ \cite{PDG}. The $\Lambda\bar{\Lambda}$ mode (branching fraction for $\chi_{c0} \simeq 0.034\%$) could also be important for spin-parity analyzing. It is also worth mentioning that the $\eta_c$ meson has a sizeable branching ratio into $p\bar{p}$ and $\Lambda\bar{\Lambda}$ (each are about 0.1$\%$ \cite{PDG}). We note that all the production processes described in this study, including $\chi_{c,b}(J)$ and $\eta_{c,b}$ CEP with the $\chi/\eta$ decaying to two fermions or two scalar particles, are incorporated in a new version of the SuperCHIC Monte Carlo \cite{SuperCHIC}.

A  rich CEP physics programme could be realised with the LHCb experiment \cite{LHCb} by employing FSC scintillation counters \cite{jerry}
surrounding the beam pipes. The excellent particle identification (in particular, the $\pi/ K$ separation) of the LHCb detector 
and the high momentum resolution are especially beneficial for measurements of the low-multiplicity decays of heavy quarkonia
or exotic states, such as the decays to the $\pi\pi$ or $K\bar{K}$ final states. An interesting study of CEP processes could be performed
with the ALICE experiment, with the diffractive double gap trigger obtained using additional detectors located on both sides of the ALICE central barrel \cite{rainer}.

Finally, we note that the CLEO  Collaboration \cite{cleo3} has recently established a significant rate of inclusive decays of the $\chi_{b}(nP)$ states to open charm with $c\bar{c}X$ accounting for about one-quarter of all hadronic decays. Further to the original prediction  of Ref.~\cite{Barbieri:1979gg},
these signals were associated predominantly with the inclusive decays of the $\chi_{b1} (1P,2P)$ bottomonium states to open charm, though the decays of the $J=0,2$ states to open charm may still be relevant. The CLEO results are in agreement with the more recent NRQCD prediction of Ref.~\cite{bodwin}. Despite the  expected suppression of the CEP $\chi_{b1}$ rate, and a smaller expected  branching fractions for $\chi_{b0,2}\to DX$ transitions \cite{Barbieri:1979gg}, it may be possible to search for $\chi_b\to DX$ CEP with the LHCb detector by exploiting its excellent vertex and proper time resolution and using the FSC system \cite{jerry} for selecting events with large rapidity gaps. We emphasise that due to the $J_z=0$ selection rule the non-resonant background to open charm production, caused by the subprocess $gg \to c\bar{c} $, is suppressed by a factor $\sim m_c^2 /M_{\chi_b}^2$. We also note that a search could be performed by LHCb for the $\chi_{c2} (2P)$ resonance signal, which has recently been observed in the $D\bar{D}$ mode \cite{belle}.

\section*{Acknowledgements}

We thank  Erik Brucken, Wlodek Guryn, Alan Martin,  Risto Orava, Jim Pinfold, Oleg Teryaev, Rainer Schicker, Antoni Szczurek
and especially Mike Albrow, for useful and encouraging discussions.
MGR, LHL and WJS
thank the IPPP at the University of Durham for hospitality. This work was supported by the grant RFBR
10-02-00040-a, by the Federal Program of the Russian State RSGSS-3628.2008.2. LHL acknowledges financial support from the University of Cambridge Domestic Research Studentship scheme.

\appendix
\renewcommand{\theequation}{A.\arabic{equation}}
\section{Proton angular distributions}\label{distap}
In this appendix we examine the expected distribution in $\phi$, the azimuthal angle between the outgoing protons, for central exclusive $\chi_{(c,b)}$ and $\eta_{(c,b)}$ production. 

\subsection{$\chi_0$}
The $\chi_0$ amplitude, given by (\ref{V0}) and (\ref{bt}), is flat in $\phi$ in the $p_\perp \ll {\bf Q}_\perp^2$ limit
\begin{equation}
A_{0^+} \propto \int\frac{{\rm d}^2 Q_\perp}{q_{1_\perp}^2q_{2_\perp}^2 Q_\perp^2}(q_{1_\perp}q_{2_\perp})\sim \int\frac{{\rm d}^2 Q_\perp}{Q_\perp^4}={\rm const}(\phi)\;,
\end{equation}
where the $q_{i_\perp}$ are defined in (\ref{qperpdef}). However, as discussed in Section~\ref{surveff}, this will receive corrections of order $\sim p_\perp^2/\langle Q_\perp^2\rangle$, resulting in some $\phi$ dependence as in Fig.~\ref{surv1}.
\subsection{$\chi_1$}
The $\chi_1$ amplitude is given by
\begin{equation}\label{A1}
A_{1^+} \propto \int\frac{{\rm d}^2 Q_\perp}{Q_\perp^2}\bigg(\frac{(q_{1_\perp})_\mu}{q_{1_\perp}^2}-\frac{(q_{2_\perp})_\mu}{q_{2_\perp}^2}\bigg)\epsilon_\chi^{*\mu}\;.
\end{equation}
Considering first the $q_{1_\perp}$ term, the integral we need to perform is therefore
\begin{equation}\label{1int}
\int{\rm d}^2 Q_\perp \frac{(Q_\perp-p_{1_\perp})_\mu}{Q_\perp^2(Q_\perp-p_{1_\perp})^2}=\int{\rm d}^2 Q_\perp \frac{(Q_\perp)_\mu}{(Q_\perp+p_{1_\perp})^2Q_\perp^2}\sim -(p_{1_\perp})_\mu  \int\frac{{\rm d}^2 Q_\perp}{Q_\perp^4}\;,
\end{equation}
where we have performed the change of variables $Q_\perp \to Q_\perp+p_{1_\perp}$ in the first step (this leaves the IR cutoffs on $|Q_\perp|$ and $|(Q_\perp-p_{1_\perp})|$ unchanged) and expanded in $(p_{1_\perp})_\mu/|Q_\perp|$ in the second. The same argument applies for the $q_{2_\perp}$ term of (\ref{A1}), giving
\begin{equation}\label{A1a}
A_{1^+} \sim (p_{2_\perp}-p_{1_\perp})_\mu \epsilon_\chi^{*\mu}\;,
\end{equation}
consistently with (\ref{R1p}) for the spin-summed amplitude squared. Clearly, the above argument is only legitimate in the strict $p_\perp \ll {\bf Q}_\perp^2$ limit and, moreover, we assume in the last step of (\ref{1int}) that the integral extends over all $Q_\perp$ when in reality we impose an IR cutoff that depends itself on $p_{1_\perp}$. This therefore constitutes a motivation of the expected $\phi$ dependence rather than a strict proof and, as usual, (\ref{A1a}) will receive corrections of order $\sim p_\perp^2/\langle Q_\perp^2\rangle$, as in Fig.~\ref{surv1}.
\subsection{$\chi_2$}
\begin{figure}[b]
\begin{center}
\includegraphics[scale=0.55]{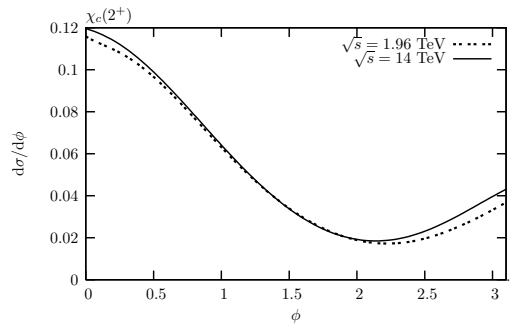}
\caption{The `bare' ${\rm d}\sigma/{\rm d}\phi$ distributions for $\chi_{c2}$ CEP at Tevatron ($\sqrt{s}=1.96$ TeV) and LHC ($\sqrt{s}=14$ TeV) energies}\label{chi2}
\end{center}
\end{figure}
The $\chi_2$ amplitude is given by
\begin{equation}\label{V2app}
A_{2^+}\propto\int\frac{{\rm d}^2 Q_\perp}{q_{1_\perp}^2q_{2_\perp}^2 Q_\perp^2}(s(q_{1_\perp})_\mu(q_{2_\perp})_\alpha+2(q_{1_\perp}q_{2_\perp})p_{1\mu}p_{2\alpha})\epsilon_\chi^{*\mu\alpha}\;.
\end{equation}
We are therefore interested in the integral
\begin{equation}\label{Int}
I_{\mu\nu}({\bf p}_{1_\perp},{\bf p}_{2_\perp})\equiv\int {\rm d}^2Q_\perp \frac{(Q_\perp - p_{1_\perp})_\mu(Q_\perp+p_{2_\perp})_\nu}{Q_\perp^2(Q_\perp - p_{1_\perp})^2(Q_\perp+p_{2_\perp})^2}\;,
\end{equation}
where $I_{\mu\nu}$ is contracted with a symmetric tensor and it is clear that $I_{\mu\nu}({\bf p}_{1_\perp},{\bf p}_{2_\perp})=I_{\mu\nu}({\bf p}_{2_\perp},{\bf p}_{1_\perp})=I_{\mu\nu}(-{\bf p}_{1_\perp},-{\bf p}_{2_\perp})$. The most general Lorentz covariant form this can then have is (omitting a $\sim g_\perp^{\mu\nu}$ term, which vanishes after the $Q_\perp$ integration, as the $\chi_2$ cannot be produced in the $J_z=0$ state in the non-relativistic quarkonium approximation, and so (\ref{V2app}) is zero for $p_\perp=0$, see Section 3 of~\cite{HarlandLang09})
\begin{equation}\label{AB}
I_{\mu\nu}=A(p_{1_\perp})_\mu(p_{2_\perp})_\nu+B\big((p_{1_\perp})_\mu(p_{1_\perp})_\nu+(p_{2_\perp})_\mu(p_{2_\perp})_\nu\big)\;,
\end{equation}
where $A$ and $B$ are in general non-trivial functions of the IR cutoff on the integral $Q_{\rm cut}$ and the ${\bf p}_\perp$, and where there is no \textit{a priori} reason to expect each of these terms not to contribute comparably to $I_{\mu\nu}$. That we have, for example, $I_{\mu\nu}({\bf p}_{1_\perp},{\bf 0})\neq 0$ and not $I_{\mu\nu}({\bf p}_{1_\perp},{\bf 0})\ll I_{\mu\nu}({\bf p}_{1_\perp},{\bf p}_{2\perp})$ for general ${\bf p}_{i_\perp}$ can be confirmed by direct integration of (\ref{V2app}). We will then have for the spin-summed amplitude squared
\begin{equation}
|A_{2^+}|^2 \sim X+Y\cos{\phi}+Z\cos^2{\phi}\;,
\end{equation}
where $X,Z>0$ and ${\rm sign}(Y)={\rm sign}(AB)$, i.e. in principle a strongly increasing or decreasing function of $\phi$, as we can see in Fig.~\ref{chi2}, which is calculated by direct numerical evaluation of (\ref{V2}).

\subsection{$\eta$}

The pseudoscalar $\eta$ amplitude is given by
\begin{equation}
A_{0^-}\propto \int\frac{{\rm d}^2 Q_\perp}{q_{1_\perp}^2q_{2_\perp}^2 Q_\perp^2} (q_{1_\perp} \times q_{2_\perp})\cdot n_0\;,
\end{equation}
where as before $n_0$ is a unit vector in the direction of the colliding hadrons (in the c.m.s. frame). As in the $\chi_2$ case, we are interested in the integral $I_{\mu\nu}$ of (\ref{Int}), but in this case contracted with the antisymmetric Levi-Civita tensor $\epsilon^{\mu\nu}$. The `B' terms of (\ref{AB}) will therefore vanish trivially, and we are left with
\begin{equation}
A_{0^-}\sim (p_{1_\perp} \times p_{2_\perp})\cdot n_0\;,
\end{equation}
consistent with (\ref{R0m}).

  S.~Heinemeyer, V.~A.~Khoze, M.~G.~Ryskin, M.~Tasevsky and G.~Weiglein,
  arXiv:0909.4665 [hep-ph].

\bibitem{BSM}
  V.~A.~Khoze, A.~D.~Martin, M.~G.~Ryskin and A.~G.~Shuvaev,
  Eur.\ Phys.\ J.\  C {\bf 68} (2010) 125
  [arXiv:1002.2857 [hep-ph]].

\bibitem{Khoze00}
  V.~A.~Khoze, A.~D.~Martin and M.~G.~Ryskin,
  Eur.\ Phys.\ J.\  C {\bf 14}, 525 (2000)
  [arXiv:hep-ph/0002072].
\bibitem{AR}
  M.~G.~Albrow and A.~Rostovtsev,
  arXiv:hep-ph/0009336.
\bibitem{KMRprosp} V.~A.~Khoze, A.~D.~Martin and M.~G.~Ryskin,
  Eur.\ Phys.\ J.\  C {\bf 23}, 311 (2002)
  [arXiv:hep-ph/0111078].
\bibitem{DKMOR}
A.~De Roeck, V.~A.~Khoze, A.~D.~Martin, R.~Orava and M.~G.~Ryskin,
  Eur.\ Phys.\ J.\  C {\bf 25}, 391 (2002)
  [arXiv:hep-ph/0207042].
%
 \bibitem{FP420}M.~G.~Albrow {\it et al.}  [FP420 R\&D Collaboration],
  JINST {\bf 4}, T10001 (2009)
  [arXiv:0806.0302 [hep-ex]].
\bibitem{bcfp} P.~Bussey, T.~Coughlin, J.~Forshaw and A.~Pilkington,
               {\em JHEP} {\bf 0611} (2006) 027 
               [arXiv:hep-ph/0607264].
\bibitem{mt1} M.~Tasevsky,
              arXiv:0910.5205.
\bibitem{royon} C.~Royon,
  Acta Phys.\ Polon.\  B {\bf 39}, 2339 (2008)
  [arXiv:0805.0261 [hep-ph]]. 
%
  \bibitem{CDFgg}
 T.~Aaltonen {\it et al.}  [CDF Collaboration],
  Phys.\ Rev.\ Lett.\  {\bf 99} (2007) 242002
  [arXiv:0707.2374 [hep-ex]].

\bibitem{CDFjj}
  T.~Aaltonen {\it et al.}  [CDF Collaboration],
  Phys.\ Rev.\  D {\bf 77},(2008) 052004 
  [arXiv:0712.0604 [hep-ex]].

\bibitem{Aaltonen09}
 T.~Aaltonen {\it et al.}  [CDF Collaboration],
  Phys.\ Rev.\ Lett.\  {\bf 102}, 242001 (2009)
  [arXiv:0902.1271 [hep-ex]].

\bibitem{Rangel} M.~Rangel [on behalf of the CDF and D0 collaborations],
talk at Rencontres de Moriond  on
QCD and High Energy Interactions,  La Thuile, March 13-20, 2010;
D0 Note 6042-CONF.

\bibitem{Albrow} Mike Albrow and Jim Pinfold, private communication.

\bibitem{Khoze04gg}
  V.~A.~Khoze, A.~D.~Martin, M.~G.~Ryskin and W.~J.~Stirling,
  Eur.\ Phys.\ J.\  C {\bf 38} (2005) 475
  [arXiv:hep-ph/0409037].

\bibitem{Pump} J. Pumplin, Phys. Rev. {\bf D47} (1993) 4820.

\bibitem{Yuan01}
 F.~Yuan,
  Phys.\ Lett.\  B {\bf 510}, 155 (2001)
  [arXiv:hep-ph/0103213].

\bibitem{petrov}
  V.~A.~Petrov and R.~A.~Ryutin,
  JHEP {\bf 0408} (2004) 013
  [arXiv:hep-ph/0403189];\\
  V.~A.~Petrov, R.~A.~Ryutin, A.~E.~Sobol and J.~P.~Guillaud,
 JHEP {\bf 0506} (2005) 007
 [arXiv:hep-ph/0409118]. 
\bibitem{bzdak}A.~Bzdak,
  Phys.\ Lett.\  B {\bf 619} (2005) 288
  [arXiv:hep-ph/0506101].

\bibitem{RPtheor} M.~Rangel, C.~Royon, G.~Alves, J.~Barreto and R.~B.~Peschanski,
  Nucl.\ Phys.\  B {\bf 774}, 53 (2007) 
  [arXiv:hep-ph/0612297]. 
%
\bibitem {Bodwin} G.~T.~Bodwin, E.~Braaten and G.~P.~Lepage,
   Phys.\ Rev.\  D {\bf 51} (1995) 1125
   [Erratum-ibid.\  D {\bf 55} (1997) 5853]
   [arXiv:hep-ph/9407339].
\bibitem{Brambilla}
 N.~Brambilla, A.~Pineda, J.~Soto and A.~Vairo,
  Rev.\ Mod.\ Phys.\  {\bf 77} (2005) 1423
  [arXiv:hep-ph/0410047];\\
 N.~Brambilla, A.~Vairo, A.~Polosa and J.~Soto,
  Nucl.\ Phys.\ Proc.\ Suppl.\  {\bf 185} (2008) 107;\\
A.~Vairo,
  arXiv:0912.4422 [hep-ph].
\bibitem{Eichten:2007qx}
  E.~Eichten, S.~Godfrey, H.~Mahlke and J.~L.~Rosner,
  Rev.\ Mod.\ Phys.\  {\bf 80}, 1161 (2008)
  [arXiv:hep-ph/0701208].
\bibitem{simon} I.~V.~Danilkin and Yu.~A.~Simonov,
  arXiv:0907.1088 [hep-ph].

\bibitem{KMRtag} V.~A.~Khoze, A.~D.~Martin and M.~G.~Ryskin,
  Eur.\ Phys.\ J.\  C {\bf 24}, 581 (2002)
  [arXiv:hep-ph/0203122].
\bibitem{herb}
S.~W.~Herb {\it et al.},
  Phys.\ Rev.\ Lett.\  {\bf 39}, 252 (1977).
\bibitem{eta}
 B.~Aubert {\it et al.}  [BABAR Collaboration],
  Phys.\ Rev.\ Lett.\  {\bf 101}, 071801 (2008)
  [Erratum-ibid.\  {\bf 102}, 029901 (2009)]
  [arXiv:0807.1086 [hep-ex]];
B.~Aubert {\it et al.}  [BABAR Collaboration],
  Phys.\ Rev.\ Lett.\  {\bf 103}, 161801 (2009)
  [arXiv:0903.1124 [hep-ex]].
\bibitem{PDG}
  C.~Amsler {\it et al.}  [Particle Data Group],
  Phys.\ Lett.\  B {\bf 667}, 1 (2008)
 and 2009 partial update for the 2010 edition.
%
\bibitem{exot}
 G.~V.~Pakhlova,
  arXiv:0810.4114 [hep-ex];\\
 G.~V.~Pakhlova,
  Phys.\ Atom.\ Nucl.\  {\bf 72} (2009) 482
  [Yad.\ Fiz.\  {\bf 72} (2009) 518];\\
  T.~Kuhr  [BaBar Collaboration and Belle Collaboration],
  arXiv:0907.4575 [hep-ex];\\
 K.~Yi,
  arXiv:0906.4996 [hep-ex].
\bibitem{frascati}
 V.~A.~Khoze, A.~D.~Martin and M.~G.~Ryskin,
  Frascati Phys.\ Ser.\  {\bf 44}, 147 (2007)
  [arXiv:0705.2314 [hep-ph]].
\bibitem{Ryskin09}
  M.~G.~Ryskin, A.~D.~Martin and V.~A.~Khoze,
  Eur.\ Phys.\ J.\  C {\bf 60} (2009) 265
  [arXiv:0812.2413 [hep-ph]].
  \bibitem{ostap}S.~Ostapchenko,
   arXiv:1003.0196 [hep-ph].

\bibitem{Guryn08}
  W.~Guryn  [STAR Collaboration],
  arXiv:0808.3961 [nucl-ex].

\bibitem{LeeDIS} J.H. Lee [on behalf of the STAR collaboration],
`Diffractive physics program with tagged forward protons at STAR/RHIC',
talk at DIS 2010, Florence, April 29-23, 2010.

\bibitem{Guryn} Wlodek Guryn, private communication.  

\bibitem{HKRSrhic} L.~A.~Harland-Lang, V.~A.~Khoze, M.~G.~Ryskin and W.~J.~Stirling, in preparation.

\bibitem{Martin01ms}
   A.~D.~Martin and M.~G.~Ryskin,
   Phys.\ Rev.\  D {\bf 64}, 094017 (2001)
   [arXiv:hep-ph/0107149].
   
   \bibitem{Shuvaev99}
  A.~G.~Shuvaev, K.~J.~Golec-Biernat, A.~D.~Martin and M.~G.~Ryskin,
  Phys.\ Rev.\  D {\bf 60}, 014015 (1999)
  [arXiv:hep-ph/9902410].

\bibitem{Coughlin09}
  T.~D.~Coughlin and J.~R.~Forshaw,
  JHEP {\bf 1001} (2010) 121
  [arXiv:0912.3280 [hep-ph]].
  
\bibitem{Martin99}
  A.~D.~Martin, R.~G.~Roberts, W.~J.~Stirling and R.~S.~Thorne,
  Eur.\ Phys.\ J.\  C {\bf 14} (2000) 133
  [arXiv:hep-ph/9907231].

\bibitem{Martin09}
  A.~D.~Martin, W.~J.~Stirling, R.~S.~Thorne and G.~Watt,
  Eur.\ Phys.\ J.\  C {\bf 63} (2009) 189
  [arXiv:0901.0002 [hep-ph]].
  
  \bibitem{MRW} A.~D.~Martin, M.~G.~Ryskin and G.~Watt,
  arXiv:0909.5529 .

  \bibitem{Kuhn79}
  J.~H.~Kuhn, J.~Kaplan and E.~G.~O.~Safiani,
  Nucl.\ Phys.\  B {\bf 157}, 125 (1979).
  
  \bibitem{Barbieri75}
   R.~Barbieri, R.~Gatto and R.~Kogerler,
   Phys.\ Lett.\  B {\bf 60}, 183 (1976).
   
\bibitem{Bergstrom79}
  L.~Bergstrom, H.~Snellman and G.~Tengstrand,
  Phys.\ Lett.\  B {\bf 82} (1979) 419.

\bibitem{Fabiano02}
  N.~Fabiano,
  Eur.\ Phys.\ J.\  C {\bf 26}, 441 (2003)
  [arXiv:hep-ph/0209283].
  
\bibitem{Parmar10}
  A.~Parmar, B.~Patel and P.~C.~Vinodkumar,
  arXiv:1001.0848.
  
\bibitem{Kwong88}
  W.~Kwong and J.~L.~Rosner,
  Phys.\ Rev.\  D {\bf 38} (1988) 279.

\bibitem{Laverty09}
  J.~T.~Laverty, S.~F.~Radford and W.~W.~Repko,
  arXiv:0901.3917 [hep-ph].
  
\bibitem{Eichten95}
  E.~J.~Eichten and C.~Quigg,
  Phys.\ Rev.\  D {\bf 52} (1995) 1726
  [arXiv:hep-ph/9503356].
\bibitem{Kim95}
  S.~Kim,
  Nucl.\ Phys.\ Proc.\ Suppl.\  {\bf 47} (1996) 437
  [arXiv:hep-lat/9510016].

  \bibitem{cball} W.~S.~Walk {\it et al.}  [Crystal Ball Collaboration],
   Phys.\ Rev.\  D {\bf 34}, 2611 (1986).
  
\bibitem{Radford07}
  S.~F.~Radford and W.~W.~Repko,
  Phys.\ Rev.\  D {\bf 75} (2007) 074031
  [arXiv:hep-ph/0701117].

  \bibitem{Barbieri:1980yp}R.~Barbieri, M.~Caffo, R.~Gatto and E.~Remiddi,
    Phys.\ Lett.\  B {\bf 95}, 93 (1980).

  \bibitem{RP}
  M.~Albrow {\it et al.},
CERN-LHCC-2006-039, CERN-LHCC-G-124, CERN-CMS-NOTE-2007-002, Dec 2006;

G.~Anelli {\it et al.}  [TOTEM Collaboration],
    JINST {\bf 3} (2008) S08007.

  \bibitem{KMRsoft}
  V.~A.~Khoze, A.~D.~Martin and M.~G.~Ryskin,
  Eur.\ Phys.\ J.\  C {\bf 18}, 167 (2000)
  [arXiv:hep-ph/0007359].
   
   \bibitem{SuperCHIC} The SuperCHIC code and documentation are available at {\tt http://projects.hepforge.org/superchic/}

\bibitem{Ryskintba}
  V.~A.~Khoze, A.~D.~Martin and M.~G.~Ryskin,	in preparation.
  
\bibitem{Gluck94}
  M.~Gluck, E.~Reya and A.~Vogt,
  Z.\ Phys.\  C {\bf 67} (1995) 433.
    
\bibitem{Peng95}
  H.~A.~Peng, Z.~M.~He and C.~S.~Ju,
  Phys.\ Lett.\  B {\bf 351}, 349 (1995).

\bibitem{Stein93}
  E.~Stein and A.~Schafer,
  Phys.\ Lett.\  B {\bf 300}, 400 (1993).

  \bibitem{Aaltonen09chi}
 T.~Aaltonen {\it et al.}  [CDF Collaboration],
  Phys.\ Rev.\ Lett.\  {\bf 102}, 242001 (2009)
  [arXiv:0902.1271 [hep-ex]].

\bibitem{Bern:2001dg}
  Z.~Bern, A.~De Freitas, L.~J.~Dixon, A.~Ghinculov and H.~L.~Wong,
  JHEP {\bf 0111} (2001) 031
  [arXiv:hep-ph/0109079].

\bibitem{Harlandlangfut}
  L.~A.~Harland-Lang, V.~A.~Khoze, M.~G.~Ryskin and W.~J.~Stirling, in preparation.

\bibitem{Brodsky81}
  S.~J.~Brodsky and G.~P.~Lepage,
  Phys.\ Rev.\  D {\bf 24} (1981) 1808.

\bibitem{Chernyak06}
  V.~L.~Chernyak,
  Phys.\ Lett.\  B {\bf 640}, 246 (2006)
  [arXiv:hep-ph/0605072].
  
\bibitem{Albrowpriv1} Mike Albrow, private communication.
  
  \bibitem{FSC} M.~Albrow {\it et al.}  [USCMS Collaboration],
   JINST {\bf 4}, P10001 (2009)
   [arXiv:0811.0120 [hep-ex]].

\bibitem{jerry}J.~W.~Lamsa and R.~Orava,
   JINST {\bf 4} (2009) P11019
   [arXiv:0907.3847 [physics.acc-ph]].

\bibitem{LHCb}A.~A.~Alves {\it et al.}  [LHCb Collaboration],
   JINST {\bf 3} (2008) S08005.

\bibitem{rainer}R.~Schicker,
   AIP Conf.\ Proc.\  {\bf 1105}, 136 (2009)
   [arXiv:0812.3123 [hep-ex]].

   \bibitem{cleo3} R.~A.~Briere {\it et al.}  [CLEO Collaboration],
   Phys.\ Rev.\  D {\bf 78}, 092007 (2008)
   [arXiv:0807.3757 [hep-ex]].

\bibitem{Barbieri:1979gg}
   R.~Barbieri, M.~Caffo and E.~Remiddi,
   Phys.\ Lett.\  B {\bf 83} (1979) 345.

\bibitem{bodwin}
  G.~T.~Bodwin, E.~Braaten, D.~Kang and J.~Lee,
   Phys.\ Rev.\  D {\bf 76}, 054001 (2007)
   [arXiv:0704.2599 [hep-ph]].

\bibitem{belle} S.~Uehara {\it et al.}  [Belle Collaboration],
  Phys.\ Rev.\ Lett.\  {\bf 96}, 082003 (2006)
  [arXiv:hep-ex/0512035];
  
  B.~Aubert  [The BABAR Collaboration],
  arXiv:1002.0281 [hep-ex].

  \end{document}